\documentclass[conference]{IEEEtran}

\usepackage{booktabs}
\usepackage{xcolor, soul}
\usepackage{xspace}
\usepackage[noend]{algorithmic}
\usepackage[algoruled, vlined, noend, scleft]{algorithm2e}
\usepackage{subfig}
\usepackage{multirow}
\usepackage{enumitem}
\usepackage{cite}
\usepackage{amsmath,amssymb,amsfonts}
\usepackage{graphicx}
\usepackage{textcomp}
\usepackage{url}
 \usepackage{flushend}

\newcommand{\hmcl}{HipMCL\xspace}
\newcommand{\bhsparse}{\texttt{bhsparse}\xspace}
\newcommand{\nsparse}{\texttt{nsparse}\xspace}
\newcommand{\rmerge}{\texttt{rmerge2}\xspace}
\newcommand{\hybrid}{\texttt{hybrid}\xspace}
\newcommand{\chash}{\texttt{cpu-hash}\xspace}

\newcommand{\isom}{\texttt{isom100-3}\xspace}
\newcommand{\isomb}{\texttt{isom100-1}\xspace}
\newcommand{\isomorig}{\texttt{isom100}\xspace}
\newcommand{\compfac}{\mathit{cf}}
\newcommand{\flops}{\mathit{flops}}
\newcommand{\nnz}{\mathit{nnz}}

\def\BibTeX{{\rm B\kern-.05em{\sc i\kern-.025em b}\kern-.08em
    T\kern-.1667em\lower.7ex\hbox{E}\kern-.125emX}}
\begin{document}

\title{Optimizing High Performance Markov Clustering for Pre-Exascale Architectures}

\author{\IEEEauthorblockN{
    Oguz Selvitopi\IEEEauthorrefmark{1},
    Md Taufique Hussain\IEEEauthorrefmark{2},
    Ariful Azad\IEEEauthorrefmark{3} and
    Ayd\i n Bulu\c{c}\IEEEauthorrefmark{4}}  
  \IEEEauthorblockA{\IEEEauthorrefmark{1}
    Lawrence Berkeley National Laboratory, Berkeley, CA (roselvitopi@lbl.gov)}
  \IEEEauthorblockA{\IEEEauthorrefmark{2}
    Indiana University, Bloomington, IN (mth@indiana.edu)}
  \IEEEauthorblockA{\IEEEauthorrefmark{3}
    Indiana University, Bloomington, IN (azad@iu.edu)}
  \IEEEauthorblockA{\IEEEauthorrefmark{4}
    Lawrence Berkeley National Laboratory, Berkeley, CA (abuluc@lbl.gov)}
}


\maketitle

%
\begin{abstract}
%
HipMCL is a high-performance distributed memory implementation of the popular
Markov Cluster Algorithm (MCL) and can cluster large-scale networks within hours
using a few thousand CPU-equipped nodes.
It relies on sparse matrix computations and heavily makes use of the sparse
matrix-sparse matrix multiplication kernel (SpGEMM).
The existing parallel algorithms in HipMCL are not scalable to Exascale
architectures, both due to their communication costs dominating the runtime at
large concurrencies and also due to their inability to take advantage of
accelerators that are increasingly popular.

In this work, we systematically remove scalability and performance bottlenecks
of HipMCL.
We enable GPUs by performing the expensive expansion phase of the MCL algorithm
on GPU.
We propose a CPU-GPU joint distributed SpGEMM algorithm called pipelined Sparse
SUMMA and integrate a probabilistic memory requirement estimator that is fast
and accurate.
We develop a new merging algorithm for the incremental processing of partial
results produced by the GPUs, which improves the overlap efficiency and the peak
memory usage.
We also integrate a recent and faster algorithm for performing SpGEMM on CPUs.
%
We validate our new algorithms and optimizations with extensive evaluations.
With the enabling of the GPUs and integration of new algorithms, HipMCL is up to
12.4x faster, being able to cluster a network with 70 million proteins and 68
billion connections just under 15 minutes using 1024 nodes of ORNL's Summit supercomputer.
  


\end{abstract}

%
%


%

%
\maketitle

\section{Introduction}
Clustering is one of the most studied problems in Computer Science, with numerous algorithms targeting different types of data and problem domains.
One of the most popular algorithms for clustering biological data, especially protein sequence similarity and protein interaction data, is the Markov Cluster (MCL) algorithm~\cite{Dongen2000}. Due to its relatively non-parametric nature and its output quality, MCL is the de-facto algorithm for clustering these datasets. However, its limited scalability makes it impossible to run on very large datasets and forces scientists to look for alternative algorithms that output lower quality clusters.

The MCL algorithm has been parallelized for multi-core and many-core
systems~\cite{Niu2014, Bustamam2012}, however the current scale of the
biological networks can make it imperative to use distributed memory systems.
In this direction, a recent study~\cite{Azad2018} proposed the \hmcl algorithm
for fast clustering of large-scale networks on distributed memory systems. 
The HipMCL algorithm returns identical clusters to MCL up to minor discrepancies due to floating-point rounding errors, but can run on thousands of compute nodes and hundreds of thousands of cores. This has opened up new opportunities for biologists to cluster their very large datasets. However, HipMCL's single-node performance is on-par with the original MCL performance so there has been little incentive for the biologists to switch to a different implementation except to take advantage of distributed memory clusters. Another downside of HipMCL was its inability to utilize GPUs, which are becoming more common in HPC centers.
This paper addresses these limitations by developing faster computational kernels, avoiding redundant computations, and overlapping communication with computation whenever possible.

Markov clustering is an iterative algorithm that operates on a graph where the edge weights in the graph represent similarity scores.
It relies on the observation that a random walk is more likely to stay in a
cluster rather than travel across the clusters.
The MCL algorithm iteratively alternates between two successive steps of
expansion and inflation until it converges.
The expansion step performs random walks of higher lengths and it enables
connecting to different regions in the graph.
The inflation step aims to strengthen the intra-cluster connections and weaken
the inter-cluster connections.  One step of the random walk from all vertices can efficiently be implemented as sparse matrix squaring, a special case of sparse matrix-matrix multiplication (SpGEMM). 

In this work, we identify various opportunities to improve the performance of HipMCL and make it ready for Exascale systems. 
To this end, we observed three performance bottlenecks in HipMCL~\cite{Azad2018} that are algorithmically solved in this paper.
First, in-node SpGEMM contributes the most to HipMCL's computation time.
Existing HipMCL uses a heap (priority-queue) assisted column-by-column algorithm for SpGEMM, which is appropriate for sparse graph processing ($\approx10$ nonzeros per column), but it becomes inefficient for relatively dense matrices in MCL ($\approx1000$ nonzeros per column).
Recent advances in SpGEMM indicate that a hash-table assisted column-by-column method is more appropriate for this setting, but care must be taken to find an optimal hybrid combination of these two methods.
We carefully benchmark the candidates that can be integrated into the HipMCL code, and find the density regimes on which each implementation dominates. Based on this recipe, we dynamically choose the best algorithm for each HipMCL iteration.



Second, HipMCL uses a bulk synchronous programming model, where processes synchronize after every major algorithmic steps. 
Here, we reduce the number of synchronization points by overlapping various operations.  
Especially, the use of GPUs opens up opportunities for overlapping in two ways: (1) we overlap the local SpGEMM computations with the MPI broadcasts within distributed SpGEMM, and (2) we overlap the CPU merging of intermediate SpGEMM products with local SpGEMM operations on the GPU. 
The latter optimization is realized by a novel merging algorithm that is completely different from the method used in existing HipMCL. 

Finally, HipMCL uses a two-pass algorithm for SpGEMM: one for the symbolic multiplication to estimate memory needs without materializing the output, and one for the actual multiplication. This almost doubles the computational time. 
We develop a sampling-based strategy that can accurately estimate the memory required for HipMCL iterations at a fraction of the cost of performing a symbolic SpGEMM. 

These three optimizations together with efficient implementations made HipMCL an order of magnitude faster than the previous implementation.
Hence, this work moved biological clustering one step closer to exascale computing, which will enable unprecedented scientific discoveries from massive-scale biological networks. Our 
optimized and GPU-enabled HipMCL code is available from the main HipMCL repository (\url{https://bitbucket.org/azadcse/hipmcl}).

%
%
%
%
%
%
%


\section{Background and Related Work}
\label{sec:bg}

%
%
Markov clustering is based on the idea of simulating flows and it performs
random walks on a given graph in order to cluster its nodes.
The algorithm performs simultaneous random walks from every vertex in each iteration, 
followed by various inflation and pruning strategies to ensure convergence and to maintain sparsity. 
While there are faster algorithms that find clusters around a set of seed vertices by only performing random walks around those seeds~\cite{spielman2013local}, such algorithms do not find the global set of clusters, unlike Markov clustering. 

%
%
%
%

%


\setlength{\textfloatsep}{1pt}

\SetAlFnt{\small}
\begin{algorithm}[t]
\caption{Overview of the MCL algorithm}
\label{alg:mcl-overview-orig}

\SetKwInput{KwIn}{Input}
\SetAlgoNoLine
\SetCommentSty{textrm}
\SetKwComment{tcc}{$\rhd$ }{}
\DontPrintSemicolon
\KwIn {A weighted network $G$}
\vspace{0.5em}

\nl  $A \gets $ Column stochastic version of the weighted adjacency matrix of $G$\;
\nl \While{change in successive iterations}{

  \nl $B \gets AA$  \tcp*{Expansion: random walks from all vertices performed as SpGEMM}

  \nl $C \gets Prune(B)$ \tcp*{Sparsify the matrix by pruning small entries (based on a user supplied threshold) and keep top-$k$ entries in every column}
  \nl $A \gets C \odot C$ \tcp*{Inflation: strengthen
the intra-cluster connections and weaken the inter-cluster connections by taking Hadamard power
of the matrix}
}
\nl \KwRet{the connected components of $A$} \;
\end{algorithm}

{\bf Notations used in this paper.}
$A_{*j}$ denotes the $j$th column of $A$ and $\mathit{inds}(A_{*j})$ denotes the set of indices of the nonzeros within that $j$th column. Let $\nnz(\cdot)$ be a function
that denotes the number of nonzero elements in a (sub)matrix.
The number of nontrivial floating point operations (i.e. $a_{ik} * b_{kj}$ where both $a_{ik}$ and $b_{kj}$ are nonzero) during the formation of the product $AB$ is denoted by $\flops(AB)$ and is
equal to
$\sum_j \sum_{i \in \mathit{inds}(B_{*j})} \nnz(A_{*i})$.
When the operands are clear from the context, we often drop the parenthesis and just use $\flops$.
The compression factor of $AB$, denoted by $\compfac(AB)$, is given by $\flops(AB) / \nnz(AB)$.
The data structures used in SpGEMM can be sensitive to the compression factor.
We occasionally omit the inputs to these two functions and use $\flops$ and $\compfac$
when it is clear from the context which matrices are multiplied.

{\bf Overview of the MCL algorithm.} Algorithm~\ref{alg:mcl-overview-orig} provides a high-level description of the MCL algorithm.
The algorithm starts with a column stochastic matrix where each column sums to 1.
In each iteration, MCL performs three successive operations (a) expansion, (b) pruning and (c) inflation.
The expansion step performs random walks from all vertices and is implemented by SpGEMM.
To keep the expanded matrix sparse, the algorithm prunes small entries based on a user supplied threshold.
If the pruned matrix is too sparse or too dense, the algorithm maintains a balanced sparsity by keeping top-$k$ entries ($k$ is typically $\sim$1000) in every column of the matrix.
Finally, the inflation step strengthens the intra-cluster connections and weakens the inter-cluster connections by taking Hadamard power
of the matrix.
When the algorithm converges, connected components of the final graph give the final set of clusters.

\begin{figure}[t]
  \centering
  \includegraphics[width=0.44\textwidth]{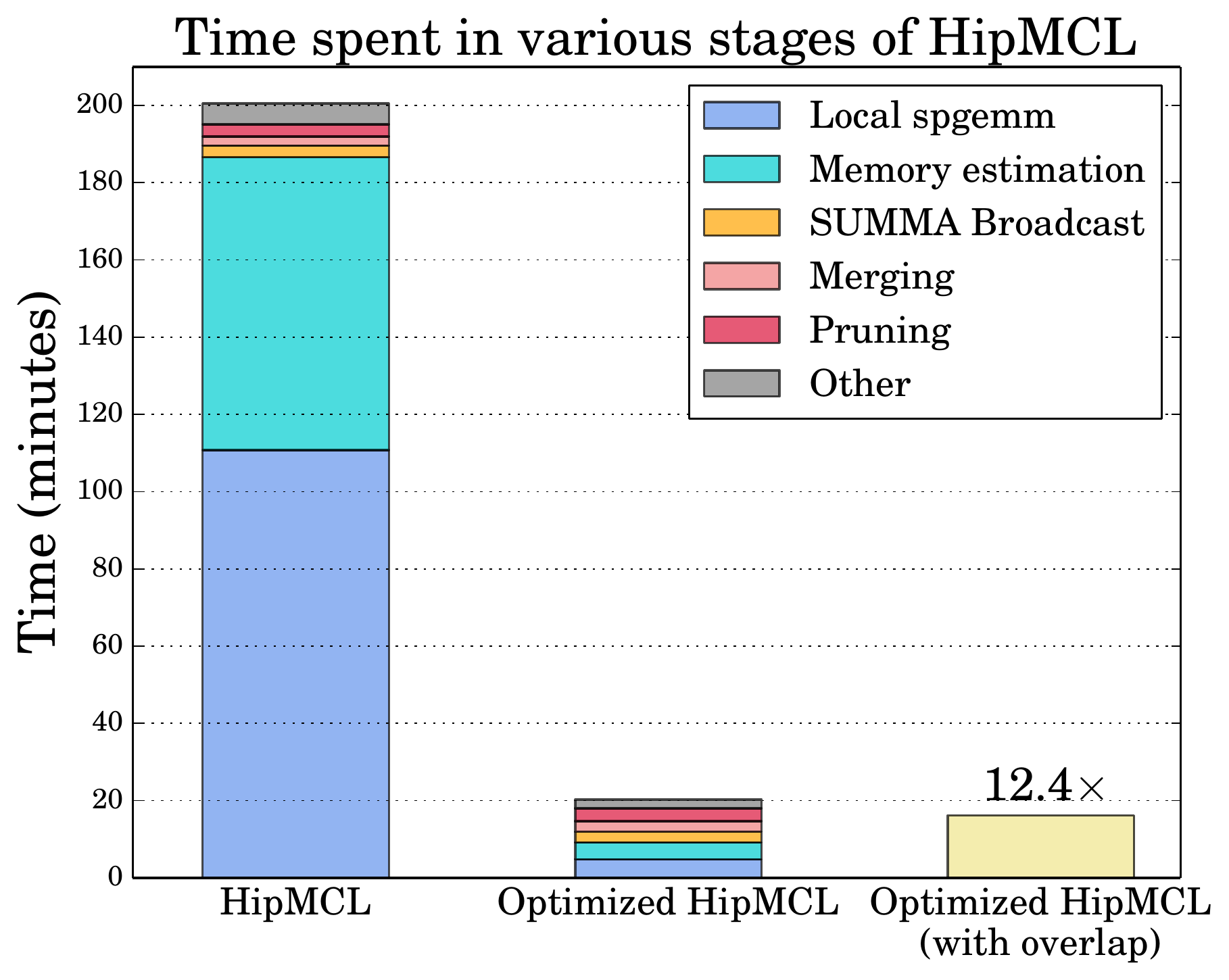}
  \caption{The running time of \hmcl and the optimized \hmcl on a network with
    35 million proteins and 17 billion connections on 100 nodes of Summit.}  
  \label{fig:dissection}
\end{figure}

{\bf Overview of HipMCL.} \hmcl is a distributed-memory parallel implementation of MCL developed using parallel sparse matrix operations. 
The expansion step is parallelized using the Scalable Universal Matrix Multiplication (SUMMA) algorithm~\cite{Geijn1995} tailored for
sparse matrices~\cite{Buluc2012} (described in the next subsection).
Parallelizing threshold-based pruning is trivial as it is performed independently in local submatrices.
HipMCL identifies top-$k$ entries in every column by selecting top-$k$ entries in each process and then exchanging these entries with other processes.
The inflation step is trivially parallelized by squaring each non-zero entry of the matrix independently.

HipMCL performs  one more optimization to handle the high memory demands of clustering of large networks.
When the unpruned matrix after expansion does not fit in the aggregate memory across all nodes, HipMCL fuses expansion and pruning and performs these two steps in several phases.
In every phase, HipMCL multiplies the first matrix $A$ by a subset of columns of the second matrix $B$ and prunes the resultant submatrix of the output.
In this way,  HipMCL avoids storing the entire unpruned matrix which may require significantly more memory than the pruned version of the output.
In order to estimate the number of phases that will be utilized in an iteration, \hmcl needs to estimate the memory needed throughout the iteration.
This estimation in \hmcl is performed by a symbolic distributed SpGEMM prior to each iteration.
Details of these parallel algorithms can be found in the original HipMCL paper~\cite{Azad2018}.

{\bf Overview of Sparse SUMMA.} 
The distributed SpGEMM implementation of HipMCL uses the Sparse SUMMA
algorithm~\cite{Buluc2012}.
This algorithm is a 2D matrix multiplication algorithm and is intensive in
subcommunicator broadcast operations.
While alternative algorithms with better bounds are known in the
literature~\cite{Ballard2013}, they require a 3D data distribution which would
limit the processor configurations that HipMCL could run.
Furthermore, the cost of redistributing the data for 3D SpGEMM is unlikely to be
amortized in the sparse case.
%

Assume that the input matrices $A$ and $B$ are both decomposed into $\sqrt{P}
\times \sqrt{P}$ blocks, where $P$ denotes the number of processes.
Let $A_{ij}$ denote the submatrix at the intersection of $i$th row stripe and
$j$th column stripe of $A$ and let $p_{ij}$ be the process responsible for
storing $A_{ij}$.
In $C=AB$, the Sparse SUMMA algorithm in stage $k$ starts with each $p_{ik}$
broadcasting $A_{ik}$ to the processes that are in the same row with $p_{ik}$
and each $p_{kj}$ broadcasting $B_{kj}$ to the processes that are in the same
column with $p_{kj}$, for $1 \leq i, j \leq \sqrt{P}$.
This is followed by the local multiplication $A_{ik} B_{kj}$.
Therefore, the task of $p_{ij}$ is to perform
$C_{ij} = \sum_{k=1}^{\sqrt{P}}A_{ik} B_{kj}$.
The summation is implemented via merging the list of intermediate products resulting from submatrix multiplications of the form $A_{ik} B_{kj}$.

{\bf Performance of different steps of HipMCL.}
Among all steps in HipMCL, expansion is the most expensive step in term of computation and communication complexity and also in practice~\cite{Azad2018}. 
 Figure~\ref{fig:dissection} illustrates the percentage of time spent in different
steps of \hmcl for a network of 35 million proteins and 17 billion connections
that is executed on 100 nodes of Summit at Oak Ridge National Laboratory.
This is also one of the networks evaluated in our experiments (\isomb).
As seen in the leftmost bar of  Figure~\ref{fig:dissection}, different steps of the expansion operations are among the most expensive operations in the original HipMCL.
Especially, the local SpGEMM computations and the memory estimation consumes  $\sim$90\% of the overall time.
In this work, we aim to improve different steps of expansion, which drastically reduces the overall runtime of HipMCL with a factor of $12.4\times$ as can be seen in rightmost bar of  Figure~\ref{fig:dissection}.
Although the overhead of communication and other stages such as merge seems to
be low, when we reduce the runtime of local SpGEMM and the memory estimation,
their overhead become comparable and this opens up several optimization
opportunities that are discussed in our study.
Therefore, the primary focus of this paper is to significantly improve the performance of the expansion step, which translates into an order of magnitude performance improvement of HipMCL.  

%
%


%
%
%
{\bf Related work on parallel SpGEMM.}
SpGEMM algorithms are extensively studied in the literature, with several parallel algorithms available for distributed memory systems~\cite{Buluc2012, Azad2016, Ballard2015, Akbudak2018},
for GPUs~\cite{rupp2016viennacl, Liu2014, Deveci2017, Nagasaka2017, Gremse2018, Dalton2015, Anh2016, kunchum2017improving}, and for multi-core
systems~\cite{Patwary2015, rupp2016viennacl, Nagasaka2018}.
Multiplying sparse matrices in parallel can be challenging for a number of
reasons.
Apart from the issues related to irregular memory accesses that are generally
associated with computations on sparse matrices, SpGEMM is especially difficult
because the nonzero pattern of the output matrix is not known in advance.
In addition, the load imbalance among parallel processing units needs to be
addressed when the input matrices are irregular.
Many works for the parallelization of SpGEMM base their approaches on
Gustavson's algorithm~\cite{Gustavson1978}.
This algorithm necessitates merging of intermediate results to get the final
product, which can be achieved in several different ways: sparse accumulators
(SPA)~\cite{Gilbert1992, Patwary2015}, expansion-sorting-compression method
(ESC)~\cite{Bell2012, Dalton2015}, heap-based accumulators~\cite{Liu2014,
  Azad2016}, hash tables~\cite{Deveci2017, Nagasaka2017}, and merge
sorts~\cite{Gremse2018}.

%
%
%

%
%
%

%


\section{Pipelined Sparse SUMMA}
\label{sec:pipe-summa}
There are several different sections in \hmcl which can benefit from running on
GPUs.
However, the most computationally expensive of them is the local SpGEMM, and it
easily tends to dominate other computations.
Local SpGEMM on CPU often consumes more time than the time consumed by all other computations.
Therefore, we exclusively focus on this operation to port on to the GPUs.
Rather than porting everything on GPU, we prefer a hybrid CPU-GPU approach due to
high memory requirements of the MCL algorithm.
We propose a hybrid algorithm called \emph{Pipelined Sparse SUMMA}, which
offloads the local SpGEMM operations in the Sparse SUMMA algorithm to the GPU
and meanwhile tries to keep CPU as busy as possible.
We first describe this algorithm and then we focus on issues related GPUs and
data formats.

\begin{figure}[t]
  \centering
  \includegraphics[width=0.50\textwidth]{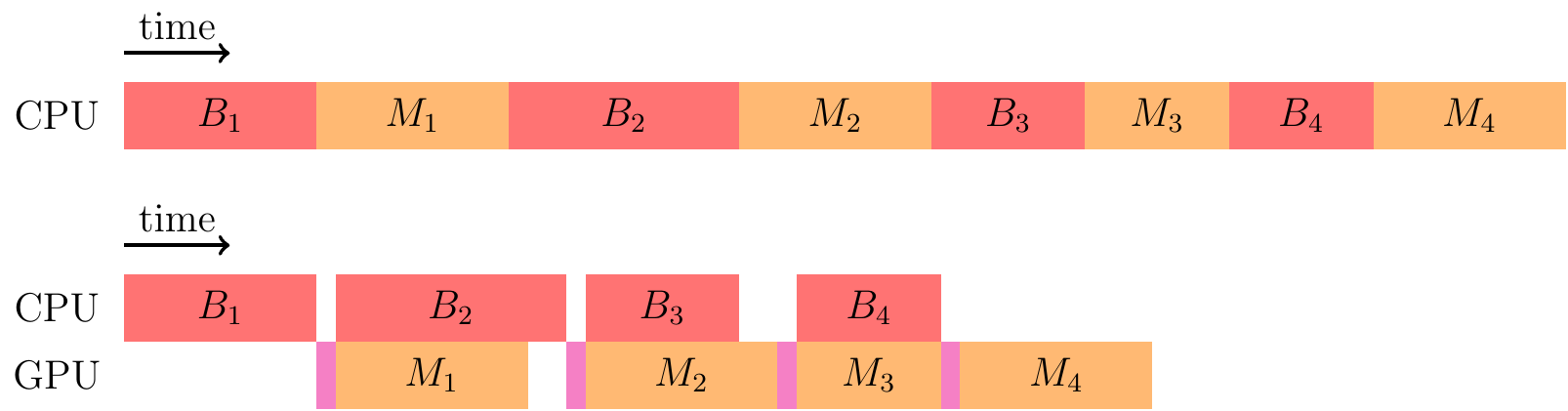}  
  \caption{A four-stage execution of the Sparse SUMMA (top) and the Pipelined
    Sparse SUMMA (bottom) algorithms. Broadcasts are indicated with $B$ and the
    multiplications are indicated with $M$. Purple boxes represent host to
    device data transfers.}
  \label{fig:overlap}
\end{figure}

HipMCL performs the expansion step in phases in order to limit the memory
consumption when it estimates that the intermediate matrix (before pruning)
cannot fit into the aggregate memory available in the system.
Performing expansion in multiple phases causes one of the input matrices to be
broadcast multiple times during the multiplication, exacerbating the
communication costs.
These broadcasts can effectively be performed by the CPU while GPU is busy with
the local SpGEMM.
The timeline for a four-stage execution of the Sparse SUMMA algorithm is
illustrated in top half of Figure~\ref{fig:overlap}, where each broadcast is
followed by a multiplication.

When local multiplication in Sparse SUMMA is performed by the GPU, the host
produces input matrices for the device via broadcasts, and the device multiplies
them and produces intermediate results for the host, which merges them.
The overlapping of multiplication in stage $k$ with the broadcast operations in
stage $k+1$ is enabled by performing local multiplication on GPU and the extra
memory provided by GPU.
As soon as the input matrices are transferred to the device, the CPU can free
the memory occupied by these matrices, and continue on to prepare the inputs of
the next stage.
In this way, the CPU only needs to wait for the transfer of the input matrices
(not the multiplication itself) to proceed into the next stage.
Hence, we can reduce the time required for the multiplication in stage $k$ and
broadcast in stage $k+1$ to maximum of those summed by the host to device
transfer time.
The timeline for a four-stage execution of the Pipelined Sparse SUMMA is
illustrated at the bottom half of Figure~\ref{fig:overlap}.
It can be seen from the figure that the closer the multiplication and the
broadcast time, the shorter the idle time for the device or the host.

Compared to the host, the device usually has limited memory.
Taking this into account, we make CPU responsible for the storage of
intermediate results and merging of them, which necessitates a fairly large
amount of memory.
The memory of GPU is only used for a single local multiplication at a time,
hence it only stores the input matrices and the respective intermediate results,
which are transferred back to the host right after the multiplication.
We describe a new and efficient merging algorithm specifically tailored for the
Pipelined Sparse SUMMA algorithm in Section~\ref{sec:lazy-merge}.

We utilize multiple libraries to perform local SpGEMM on GPU.
These libraries are: \bhsparse~\cite{Liu2014}, \nsparse~\cite{Nagasaka2017}, and
\rmerge~\cite{Gremse2018}.
There are two important criteria in selecting where (i.e., CPU or GPU) and with
which library (i.e., \bhsparse, \nsparse, or \rmerge) to perform the local
SpGEMM on GPU: $\flops$ and $\compfac$.
The $\flops$ metric is used to decide whether to perform the multiplication on
GPU or CPU.
In the case where there are not enough arithmetic operations to saturate the GPU
threads, we perform multiplication on CPU.
Otherwise, we perform SpGEMM on GPU by selecting the library for that purpose
according to $\compfac$.
%

%
%

\subsection{Managing multiple GPUs on a node}
The management of multiple GPUs on a single node can be done in several ways.
The most straightforward and effortless approach is to use a single MPI process
per GPU.
This approach might be viable for applications that does not heavily make use of
CPU resources.
However, in \hmcl we keep CPU responsible for several tasks involving
communication and multi-threaded computations while the GPU is busy with the
local SpGEMM.
Therefore, if there are $r$ GPUs and $s$ cores on a single node, the question
reduces to choosing between using $r$ MPI tasks each with $s/r$ threads and
using a single MPI task with $s$ threads for computations on CPU.
%
We found that using a single MPI process per node and relying on intra-node
multi-threading was more effective for HipMCL
(Section~{\ref{sec:eval-pipe}}).

We divide local SpGEMM computations on a single node among GPUs by following a
similar approach to the phased execution of the expansion step in \hmcl.
Local $C=AB$ is computed by entirely copying $A$ to all GPUs devices and
dividing columns of $B$ evenly among GPUs.
This results in each GPU computing its own portion of the output matrix and
makes the formation of the final output matrix trivial.

\subsection{Sparse matrix storage format}
The data structures for storing sparse matrices in \hmcl differ from those
utilized in \bhsparse, \nsparse, and \rmerge.
All these three libraries rely on compressed sparse row format (CSR) for storing
matrices.
\hmcl, on the other hand, relies on doubly compressed sparse column format
(DCSC), which is designed for efficient representation and processing of
hypersparse matrices~\cite{hypersparse08}.
DCSC format basically uses an additional array compared to the compressed sparse
column format (CSC) by also compressing the column pointers.

The conversion from DCSC to CSR format is not fully necessary and it is possible
to save from preprocessing time by making use of most of the underlying
structures already existent.
We first get the CSC format from DCSC by simply decompressing the column
pointers.
This does not alter the index and value arrays (each size of number of nonzeros)
stored in the DCSC format and just creates and fills a new array for the
decompressed column pointers.
Considering we have $A$ and $B$ stored in CSC format, the conversion from CSC to
CSR format may look necessary at first glance to perform $C=AB$.
However, storing a sparse matrix in CSC format is equivalent to storing its
transpose in CSR format.
Hence, computing $C^T = B^TA^T$ with all matrices in CSR format actually
produces the result matrix $C$ in CSC format, which is then processed within
\hmcl.
Therefore, no explicit conversion is necessary if we compute $BA$ with both
matrices in CSC format.

\section{Binary merge}
\label{sec:lazy-merge}
There is a further optimization opportunity in keeping both the GPU and the CPU
busy by performing the merging of intermediate products on CPU while the
performing local SpGEMM on GPU.
However, the existing HipMCL implementation would not allow this as it defers
the merging of intermediate products (i.e., the result of each $A_{ik} B_{kj}$)
until all stages of the Sparse SUMMA complete.
While this is asymptotically fast, it limits opportunities for hiding the cost
of merging.
To this end, we implement a binary merge scheme.

The Sparse SUMMA algorithm executes in $k = \sqrt{P}$ stages and in each stage,
a set of intermediate results are produced for the output matrix.
The intermediate results for an element of the output matrix are merged (i.e.,
summed) to get the final value.
A heap data structure of a maximum size of $k$ is utilized for merging of the
intermediate results, which results in a variant of the algorithm known as
$k$-way merge or multiway merge~\cite{Knuth1998-3}.
For the analyses and discussions in this section, we assume each of the $k$
lists to be merged contains $n$ elements and they are pairwise disjoint in terms
of the output matrix indices they contain.
The latter assumption, although does not make much difference in the runtime of
multiway merging, leads to worst-case behavior of the binary merge, and hence it
is helpful to get the upper bound.
The complexity of multiway merging is $O(kn\lg{k})$.

After local SpGEMM computation on GPU, we can start merging the intermediate
results on CPU in an \emph{incremental} manner without needing to wait for all
of the $k$ lists to do the merge.
In this way, the computations for merging can be overlapped with the local
SpGEMM computations, hence we can keep CPU busy while GPU proceeds to the next
stage of the Sparse SUMMA algorithm.
Note that the merging is often cheaper than the local SpGEMM in
\hmcl.

The merging can be done immediately when an intermediate result list becomes
available by merging it with the already existing merged results computed from
the merges in the previous stages.
This approach consists of $k-1$ successive two-way merges (i.e., no heap) and
contains a total of $n(k (k + 1)/2-1)$ operations, assuming merging of
  two lists are in linear in terms of sizes of both lists.
Although modest in term memory usage compared to multiway merge, immediately
merging results performs many redundant passes
over the intermediate results. It also continuously occupies the CPUs, which are also tasked with the 
relatively more expensive broadcast operations in the Sparse SUMMA algorithm.
Taking these issues into consideration, we propose a cheaper alternative called
\emph{binary merge}, which merges intermediate results only in even-numbered
stages of the Sparse SUMMA algorithm.
This algorithm is illustrated in Algorithm~\ref{alg:lazy-merge}.
%

\begin{algorithm}[t]
\caption{Binary merge}
\label{alg:lazy-merge}

\SetKwInput{KwIn}{Input}
\SetAlgoNoLine
\SetCommentSty{textrm}
\SetKwComment{tcc}{$\rhd$ }{}
\DontPrintSemicolon
\KwIn {number of stages ($nstages$)}
\KwOut {merged list}
\vspace{0.5em}

\nl Initialize an empty stack $S$ \;
\nl \For{$i\leftarrow 1$ \KwTo $nstages$}{
  \nl Wait for list $l_i$ to be available \;
  \nl $S.push(l_i)$ \;
  \nl $j \leftarrow i$, $nmerges \leftarrow 0$ \;
  \nl \While{$j \text{ is even } \mathbf{and} \text{ } j \neq 0$} {
    \nl $nmerges \leftarrow nmerges + 1$ \;
    \nl $j \leftarrow j / 2$
  }
  \nl \If{$nmerges = 0$} {
    \nl continue \;
  }
  \nl Let $L$ be an array of size $nmerges+1$ \;
  \nl \For{$j \leftarrow 1$ \KwTo $nmerges+1$} {
    \nl $L[j] \leftarrow S.pop()$ \;
  }
  \nl $merged\_list \leftarrow$ merge all lists in $L$ \;
  \nl $S.push(merged\_list)$ \;
}
\nl \KwRet{$S.pop()$} \;
\end{algorithm}

The binary merge algorithm has practically the same merging structure of the
merge sort algorithm.
The basic difference is that since the lists come in order as they become
available, the algorithm cannot follow a divide and conquer design.
The runtime complexity of Algorithm~\ref{alg:lazy-merge} is $O(kn\lg{k})$
assuming that the merging of the lists in line 14 is performed with successive
two-way merges.
Note that there may be more than two lists to be merged in $L$ in the algorithm.
In practice we found performing successive two-way merges inefficient, and
instead we choose to merge all the lists in $L$ by using a heap.
Assuming $k$ is a power of 2, this leads to
\begin{align*}
  \begin{split}
     &\frac{kn}{2} \sum_{i = 2}^{\lg{k}} {\lg{i}} + kn\lg{(\lg{k}+1)} 
     = \frac{kn}{2} (\sum_{i = 2}^{\lg{k}} {\lg{i}} + 2\lg{(\lg{k}+1)}) \\
     &= \frac{kn}{2} (\lg{(\lg{k} + 1)!} + \lg{(\lg{k}+1)}) 
     < \frac{kn}{2} (\lg{(\lg{k} + 2)!})
  \end{split}
\end{align*}
number of operations, which derives from the fact that the sizes of the heaps
used in the merges are different and bounded by $\lg{k}+1$.
Since $\lg{n!} = \Theta(n\lg{n})$,
\begin{align*}
  \begin{split}
    & \frac{kn}{2} (\lg{(\lg{k} + 2)!}) = \frac{kn}{2} (\lg{k}+2) (\lg(\lg{k}+2)) \\
    & = O(kn \lg{k} \cdot \lg{\lg{k}}),
  \end{split}
\end{align*}
which is only a factor of $\lg{\lg{k}}$ worse than $O(kn\lg{k})$.
%
%
In terms of computational efficiency, binary merge is slightly inferior to the
multiway merge.
However, it enables two critical optimizations.
First, we are able to overlap it with the expensive local SpGEMM.
Second, it is more efficient in terms of memory usage since the multiway merge
necessitates storing all intermediate results.
Considering some of the intermediate results in binary merge will be compressed
along the way, the final merge is likely to contain fewer intermediate results
than the multiway merge.

\section{Probabilistic memory requirement estimation}
\label{sec:sampling}
To cluster large networks, \hmcl executes the Sparse SUMMA algorithm in $h$
phases in order to not to exceed the available memory at each process.
Instead of expanding and pruning the entire matrix altogether, \hmcl expands and
prunes $b$ columns at a time in each phase, where $b$ can be adjusted to achieve
a trade-off between memory requirements and computational efficiency.
To find the number of phases, \hmcl first needs to estimate the memory required
prior to running an MCL iteration.
Note that the estimation needs to be done at the beginning of \emph{each}
iteration of the MCL algorithm as the input network constantly changes across
the iterations.
This estimation in \hmcl is done via a symbolic SpGEMM within the Sparse SUMMA
algorithm using the entire network.
It inherently necessitates less memory compared to the numeric multiplication
since the output matrix is never materialized.
However, it is expensive, i.e., $O(\flops)$, and most of the challenges related
to numeric SpGEMM still exist for the symbolic multiplication as well.
Since we are just estimating the amount of memory needed through an MCL
iteration, it is reasonable to make a probabilistic estimation, which may result
in fewer or more number of phases compared to the correct value.
This can easily be compensated by using a slightly smaller input value for the
actual available memory of each process.

\begin{figure}[t]
  \centering
  \includegraphics[width=0.48\textwidth]{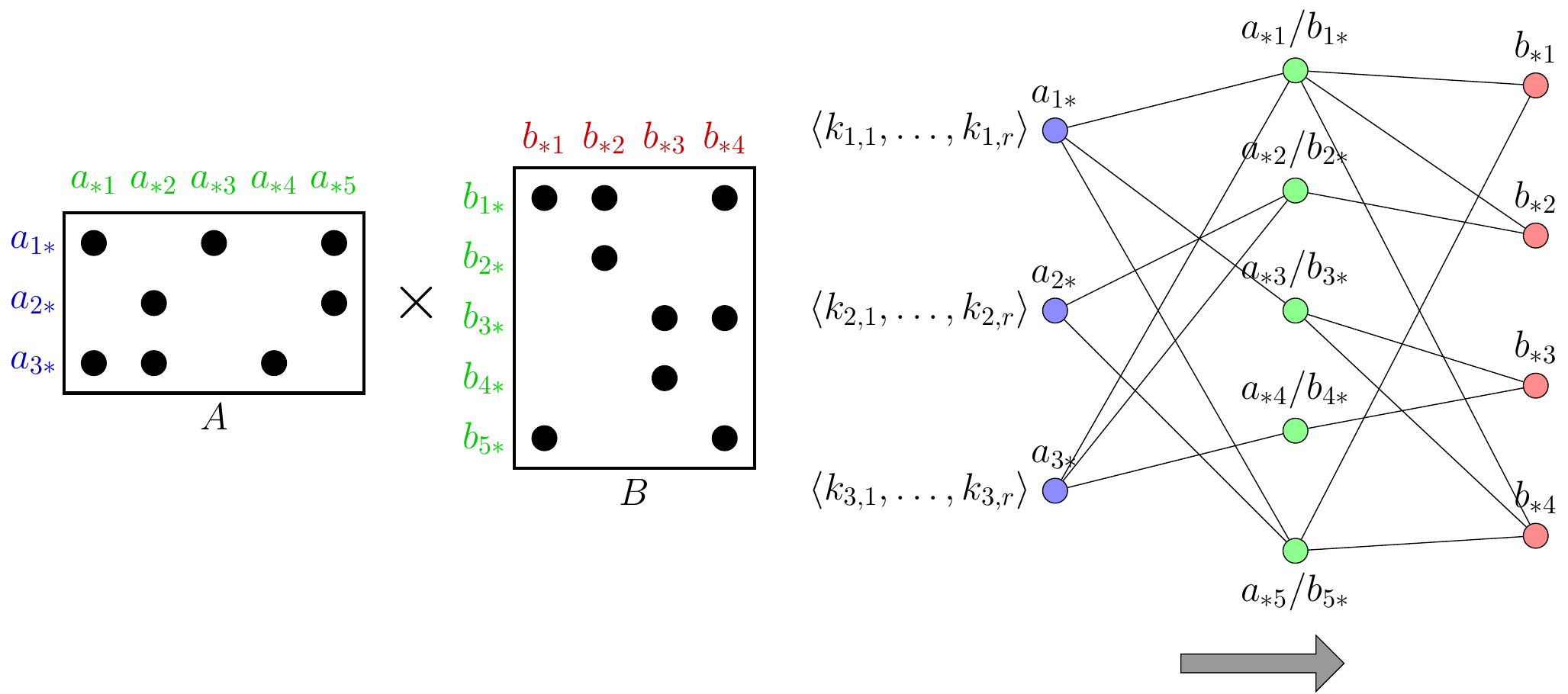}  
  \caption{The layered graph that represents an example SpGEMM instance. The
    keys associated with the vertices in the first layer are propagated across
    successive layers.}  
  \label{fig:layered-graph}
\end{figure}

To estimate the number of output elements in SpGEMM, we adapt the method
proposed by Cohen~\cite{Cohen1998}.
This method makes use of the layered structure of $C=AB$, by capturing it with a
three-layer graph, where the vertices in the first layer represent the rows of
$A$, the vertices in the middle layer represent the columns of $A$ or rows of
$B$, and the vertices in the third layer represent the columns of $B$.
Figure~\ref{fig:layered-graph} shows an example layered graph that represents
multiplication of a $3 \times 5$ and a $5 \times 4$ matrix.
It assigns random keys following an exponential distribution to the vertices in
the first layer, and then propagates the smallest keys across the layers.
The crux of this method is that for a vertex $v$ that is in any
layer other than the first layer, there is a correlation between the number of
first-layer vertices that reach $v$ and the key of the minimally ranked vertex
among them.
For each vertex in the first layer, the algorithm uses $r$ keys, $k_{i,1}
\ldots, k_{i,r}$, drawn from an exponential distribution.
After propagating the keys across layers it makes the estimation
$(r-1)/\sum_{j=1}^r k_{i,j}$ on the final keys.
The accuracy of the estimation gets better with increased number of keys.

The runtime of this probabilistic estimation method is $O(r \cdot (\nnz(A)+\nnz(B)))$.%
There are several advantages to it compared to the estimation with symbolic
SpGEMM.
First, when $\compfac$ is large, its complexity is expected to be smaller than
the $O(\flops)$ runtime complexity of the symbolic SpGEMM.
Note that the most time-consuming local SpGEMM operations in \hmcl possess a
large $\compfac$.
Although the runtime of the algorithm increases proportional to the number of
keys used, with our experiments we confirm that a small number of keys is
able to produce good estimations.
Moreover, the probabilistic estimation is easy to parallelize with a
vertex-based task distribution and the processing of each set of keys
necessitates no communication or coordination among threads or processes.
Another advantage of this method is that its memory requirements are low, which
is proportional to the number of keys and number of rows/columns of the matrices.
%

\section{Faster SpGEMM on CPU}
\label{sec:hash-spgemm}
In this section we describe the adaption of a recent
algorithm~\cite{Nagasaka2018} for performing SpGEMM on CPU.
This specifically aims at optimizing \hmcl for the systems without GPUs.
As discussed in Section~\ref{sec:bg}, the merging of the intermediate results
produced for an output column in SpGEMM can be performed with different data
structures.
\hmcl relies on heaps for this purpose.
Note that this is similar to the merging problem in
Section~\ref{sec:lazy-merge}, which contains the intermediate results produced
across different stages of the Sparse SUMMA algorithm.
Here, we solely focus on producing output results within a single iteration of
the Sparse SUMMA algorithm, i.e., single local SpGEMM.
The same structures could be used for both of these merge operations, however,
we found out heaps to be performing better in merging results produced in
different stages of the Sparse SUMMA algorithm.
Within a single SpGEMM, however, hash tables prove to be better for the reasons
described below.

Similar to their previous work addressing SpGEMM on GPUs~\cite{Nagasaka2017},
the work for CPU-based SpGEMM~\cite{Nagasaka2018} also uses hash tables, but
this time for shared-memory parallel SpGEMM on multi-core systems (Intel Xeon
and KNL).
This approach maintains a different hash table for each thread.
The size of a hash table is determined according to the maximum number of
$\flops$ per row among the rows assigned to this thread and the same hash table
is utilized throughout the thread's lifetime.
The symbolic multiplication is performed by simply inserting keys into the hash
table, while the numeric multiplication additionally necessitates the update of
the values.
The final values for a column are obtained by sorting the hash table after all
intermediate results are processed.
%

For SpGEMM instances with small $\compfac$ values, the heaps tend to perform
better than the hash tables and for the instances with large $\compfac$ values,
the hash tables outperform the heaps~\cite{Nagasaka2018}.
%
Hash tables are the recommended data structures for the SpGEMM when $\compfac$
is large and the nonzeros in the columns are sorted with respect to their row
identifiers.
%
Since the most time-consuming multiplications have a large $\compfac$ in \hmcl
as we will argue in our evaluation, hash-table-based SpGEMM is expected to make
a big difference on CPU.
%
For these reasons, we utilize hash tables for SpGEMM on CPU.

\section{Evaluation}
\label{sec:eval}
\subsection{Datasets and Setup}
\hmcl heavily makes use of the sparse matrix operations and data structures
provided by the Combinatorial BLAS library (CombBLAS)~\cite{combblas2011}.
CombBLAS is a distributed-memory parallel graph analytics library written in
C++ using MPI and OpenMP.
We have integrated GPU support for performing SpGEMM into CombBLAS and
significantly enhanced it with the described optimizations.
%
Three GPU codes \bhsparse, \nsparse, and \rmerge are made accessible to calls by
CombBLAS sparse matrix storage formats with a common interface so that they can
be used independently throughout the execution.
%
The distribution of work among the threads on a node depends on the stage of the
HipMCL algorithm being run and how it is parallelized.
There are several distinct stages, such as merging, memory requirement
estimation, selection, etc. In most of these stages, a dynamic work
distribution is utilized through OpenMP. For tasks that require more involved coordination, we use pthreads.
%
We used g++ 5.4.0 to compile the host code and Cuda Toolkit 9.2 to compile the
device code.

We conduct our experiments on the IBM supercomputer Summit hosted at Oak Ridge
National Laboratory (number one system as of writing of this paper at Top500
list~\cite{Top500}).
Summit has a total of 4608 nodes.
Each node contains two IBM Power9 CPUs (each with 22 cores clocked at 3.3 GHz
and 256 GB memory) and six NVIDIA Volta V100 GPUs (each with 80 streaming
multiprocessors and 16 GB memory).
The nodes are connected in a non-blocking fat-tree topology using a dual-rail
Mellanox EDR InfiniBand interconnect.

\begin{table}[t]
  \centering
  \caption{Three medium-scale (top half) and three large-scale networks (bottom
    half) used in the experiments.}
  \scalebox{1.00} {
    \begin{tabular}{l r r}
      \toprule
      network & \#proteins & \#connections \\
      \midrule
      \texttt{archaea} & 1,644,227 & 204,784,551 \\
      \texttt{eukarya} & 3,243,106 & 359,744,161 \\
      \isom & 8,745,542 & 1,058,120,062 \\
      \midrule
      \isomb & 35M & 17B \\
      \isomorig & 70M & 68B \\
      \texttt{metaclust50} & 383M & 37B \\
      \bottomrule
    \end{tabular}%
  }
  \label{tb:dataset}%
\end{table}%

In order to validate the proposed optimizations for HipMCL, we conduct
experiments using relatively few compute nodes using the three
medium-scale networks in top half of Table~\ref{tb:dataset}.
The \texttt{archaea}, \texttt{eukarya}, and \isom networks respectively contain
Archaeal, Eukaryotic, and all proteins from the isolate genomes in the
Integrated Microbial Genomes (IMG) database~\cite{Chen2017}.
For \isom, we reduced the size of the Isolate-3 graph (\isomorig) from the
original \hmcl paper~\cite{Azad2018} by extracting an induced subgraph whose
vertices are a random $|V|/8$-sized subset of the vertices in the original
graph.
After validating our optimizations on these three networks, we finally evaluate
the overall performance of \hmcl with all optimizations enabled on three larger
networks given in bottom half of Table~\ref{tb:dataset}.
\isomb is generated following the methodology for generating \isom by using
$|V|/2$-sized subset of the vertices in the original graph.
The \texttt{metaclust50} network contains the similarities of proteins in
Metaclust50\footnote{\url{https: //metaclust.mmseqs.com}}, which consists of the
predicted genes from metagenomes and metatranscriptomes of assembled contigs
from IMG/M and NCBI.
We use an inflation parameter of 2 for \hmcl in all experiments.

\subsection{Pipelined Sparse SUMMA}
\noindent
\textbf{Algorithm selection for local SpGEMM.}
\label{sec:eval-pipe}
We evaluate the performance of local SpGEMM computations on CPUs and GPUs for
different algorithms.
For both CPU and GPU, we rely on two metrics in selecting an algorithm for
performing SpGEMM: $flops$ and $cf$.
For local SpGEMM on CPU, we found hash-table-based algorithm, abbreviated as
\chash, to be more effective than the heap-based algorithm for most of the time.
%
For small $cf$ values, the heaps show themselves to be slightly more effective
while for large $cf$ values hash tables perform significantly better.
For local SpGEMM on GPU, among the tested three libraries \nsparse is found to
perform significantly better than \bhsparse and \rmerge for large $cf$, while
for small $cf$, \rmerge is found to be slightly better.
%
We also evaluated a hybrid scheme, abbreviated as \hybrid, which selects one of
the four schemes (\chash, \nsparse, \bhsparse, \rmerge) in multiplication
according to $\compfac$ and the $\flops$ of the SpGEMM.
%

\begin{figure}[t]
  \centering
  \includegraphics[width=0.41\textwidth]{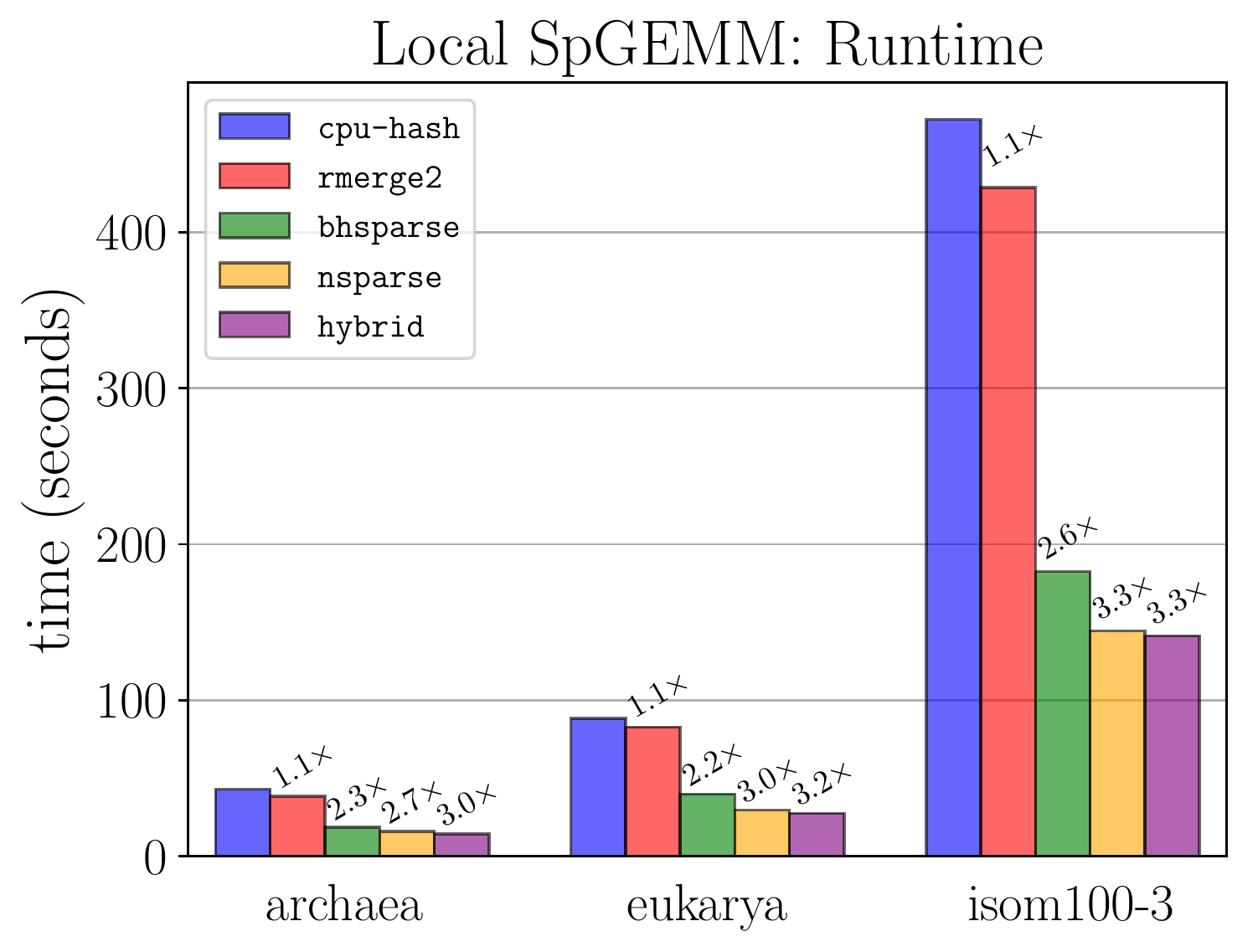}
  \caption{Overall time spent in SpGEMM.}
  \label{fig:tspgemm}
\end{figure}

Figure~\ref{fig:tspgemm} presents the time spent in SpGEMM by the tested schemes
for three different networks.
The SpGEMM greatly benefits from running on GPU: \rmerge, \bhsparse, and
\nsparse are up to 1.1$\times$, 2.6$\times$, and 3.3$\times$ faster than \chash,
respectively.
The \hybrid scheme slightly improves the SpGEMM runtime compared to the best
performing GPU library \nsparse from 2.7$\times$ to 3$\times$ for the
\texttt{archaea} network and from 3$\times$ to 3.2$\times$ for the
\texttt{eukarya} network.


\noindent
\textbf{Overlap efficiency.}
We analyze the effect of overlapping local SpGEMM computations on GPU with the
broadcasts and the merging of intermediate results on CPU.
Note that we are able to hide the overhead of either the operations on GPU or
the operations on CPU.
Ideally the time to perform these operations is given by the unit that takes
longer.
Our experimental setup for testing the effectiveness of overlapping CPU and GPU
operations consists of testing three networks on three different number of nodes
16, 36, and 64.
We measure the individual time to perform SpGEMM on GPU (including data
transfers, pre/postprocessing), to perform the broadcasts, to perform the binary
merge, and finally the actual time that takes to perform these with the
overlaps.
Table~\ref{tb:overlap} presents the obtained results.

As seen in Table~\ref{tb:overlap}, we are able to hide most of the computations
on CPU.
In the table, the computations on GPU take longer than the computations on CPU
(i.e., compare SpGEMM time with the summation of the broadcast and merge time).
Hence, the SpGEMM time determines the runtime as it takes longer.
The overall runtime is close to the SpGEMM time and it is 15\%-20\% higher than
it.
Ideally if all the time spent on CPU were hidden, these two values would have
been equal.
However, there are certain operations whose overhead we cannot hide.
For example, the first broadcast or the final merge operations (which is the
most expensive one in the binary merge) cannot be hidden because GPU is idle
during those.
Another such overhead is the host and device transfers.
These add up and cause the overall runtime to be higher than the ideal runtime.

\setlength{\textfloatsep}{3pt}

\begin{table}[t]
  \centering
  \caption{Overlap efficiency.}
  \scalebox{1.00} {
    \begin{tabular}{r r r r r r r}
      \toprule
      & & \multicolumn{4}{c}{time (seconds)} &  \\
      \cmidrule(r{4pt}){3-6} 
      network & \#nodes & SpGEMM & bcast & merge & overall  \\
      \midrule
      \multirow{3}{*}{\small \texttt{archaea}} & 16 & 14.6 & 3.4 & 3.1 & 17.2 \\
                                               & 36 & 9.5 & 4.4 & 2.4 & 11.4 \\
                                               & 64 & 7.1 & 3.4 & 1.7 & 8.4  \\
      \midrule                                        
      \multirow{3}{*}{\small \texttt{eukarya}} & 16 & 27.2 & 5.3 & 6.1 & 32.2  \\
                                               & 36 & 20.0 & 7.5 & 5.4 & 23.2  \\
                                               & 64 & 14.0 & 6.8 & 3.6 & 16.5  \\
      \midrule
      \multirow{3}{*}{\small \isom} & 16 & 145.7 & 39.9 & 26.7 & 170.5  \\
                                    & 36 & 99.9  & 28.2 & 26.0 & 114.8  \\
                                    & 64 & 55.1  & 19.2 & 15.3 & 64.8   \\
    \bottomrule
    \end{tabular}%
    }
  \label{tb:overlap}%
\end{table}%

\begin{figure}[t]
  \centering
  \subfloat{\includegraphics[width=0.245\textwidth]{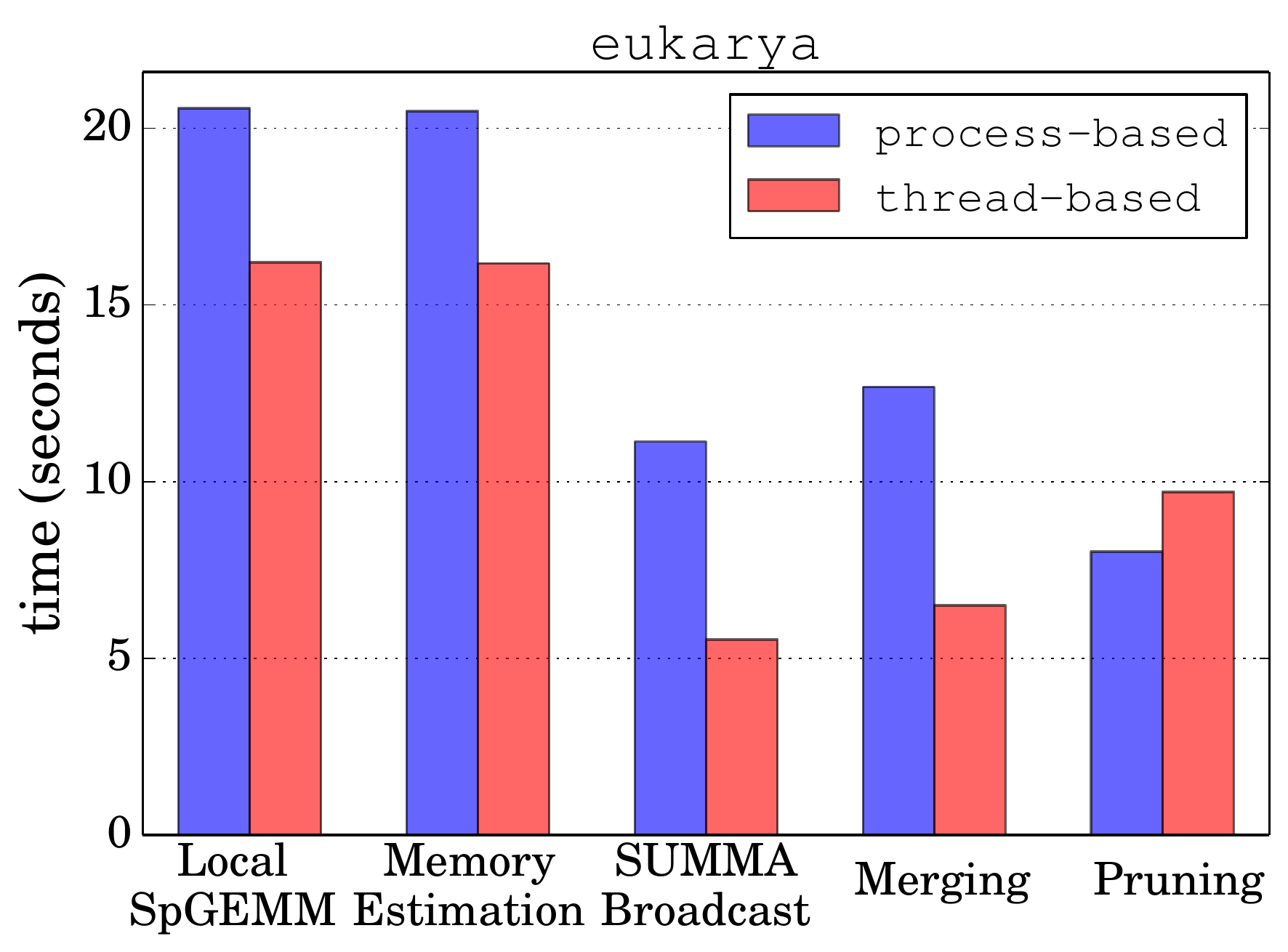}}
  \subfloat{\includegraphics[width=0.245\textwidth]{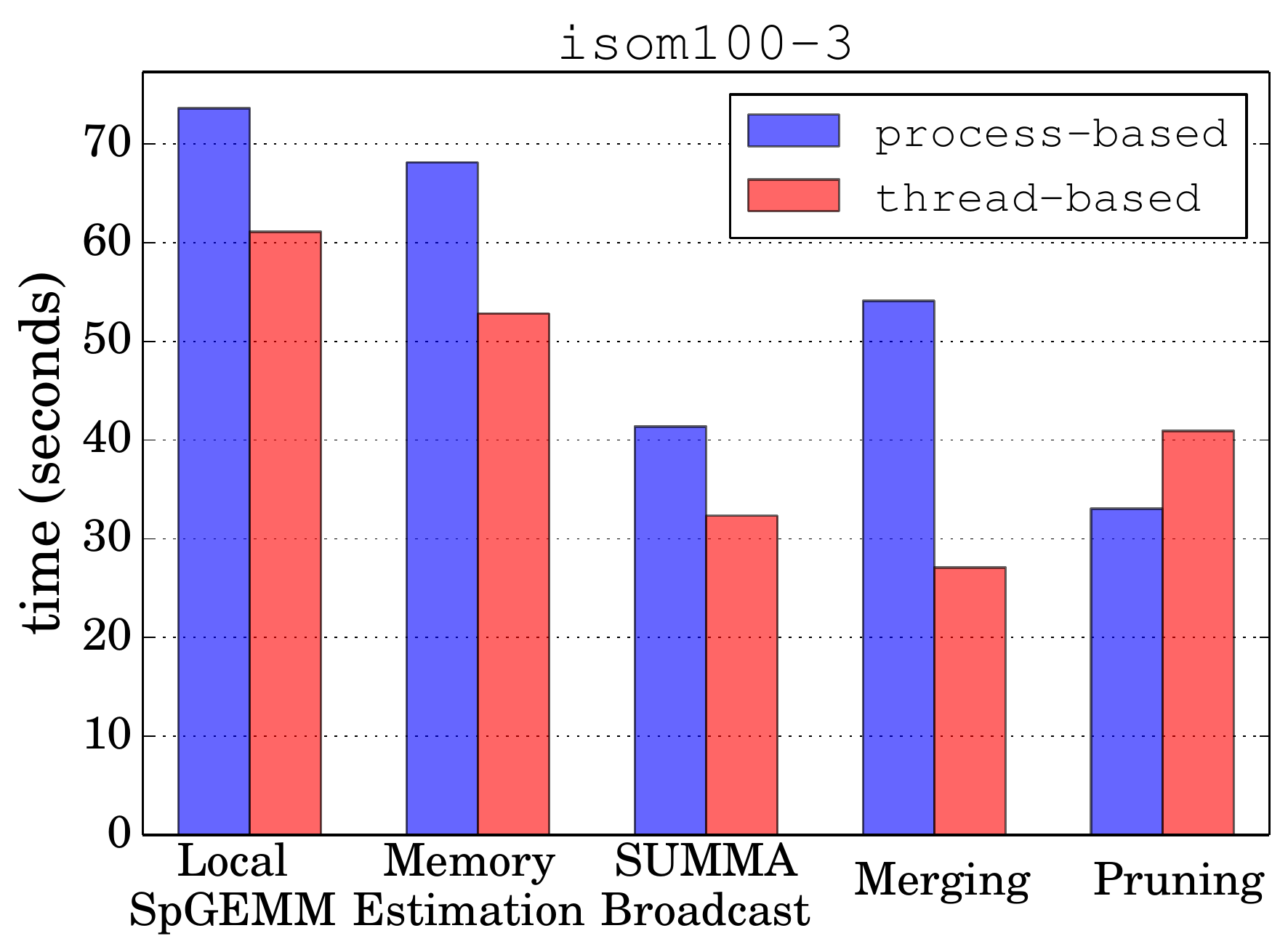}}   
  \caption{Threads vs. processes in managing a node's resources.}
  \label{fig:multigpu}
\end{figure}

\noindent
\textbf{Managing multiple GPUs on a node}
We evaluate the efficiency of controlling GPUs on a single node by comparing the
utilization of threads and MPI processes for that purpose.
In our experimental setting, we distribute the computational resources of 16
nodes among (i) 16 MPI processes, one process per node, with each process
assigned 40 threads and 4 GPUs on a node and (ii) 64 MPI processes, four
processes per node, with each process assigned 10 threads and a single GPU.
We refer to the former as the thread-based setting and the latter as the
process-based setting.
This evaluation aims to find out whether threads or processes are more effective
in controlling the resources on a node for HipMCL.
We use 4 GPUs on a node instead of all 6 GPUs due to the HipMCL's requirement of
perfect square numbers as the number of MPI processes as input.
We note this underutilization of GPUs only happens for the purposes of
evaluation of threads versus\, processes and all GPUs are properly utilized in
our final code.
We report the time spent in five different stages of HipMCL by the thread-based
and process-based settings in Figure~{\ref{fig:multigpu}} for {\texttt{eukarya}}
and {\isom} networks.
We omit the {\texttt{archaea}} network as the it exhibits similar behavior to
these two networks.

Figure~{\ref{fig:multigpu}} shows that the thread-based setting setting is more
effective than the process-based setting for all stages except the pruning stage.
For the {\isom} network, the thread-based setting is 13\%, 23\%, 19\%, and 50\%
faster than the process-based setting in the local SpGEMM, memory requirement
estimation, SUMMA broadcast, and merging stages, respectively, while it is 24\%
slower in the pruning stage.
Similar values are observed for the {\texttt{eukarya}} network.
The thread-based setting seems to be more effective in controlling GPUs and
communicating data in HipMCL.
We use the thread-based setting for our large-scale experiments in the following
section.

\subsection{Binary merge}
We assess the performance of the proposed binary merge scheme in terms its
runtime and memory requirements by comparing it to the multiway merge.
Recall that the binary merge scheme relies on GPU to perform the local SpGEMM
for overlapping them and it incrementally merges the output matrices on CPU in
the Sparse SUMMA algorithm, while the multiway merge scheme waits for the
results of the all stages of the Sparse SUMMA algorithm become available.
%
%
For the experiments in this section we used 16 nodes and 40 threads (i.e.,
cores) per node for performing the merge.
We test these two schemes on three different networks.

%
%
The binary merge and the multiway merge exhibit very similar runtime
performances.
On total, the binary merge is only 3\%, 4\%, and 3\% slower than the multiway
merge for the networks \texttt{eukarya}, \texttt{archaea}, and \isom
respectively.
These values are in line with the runtime complexities of these two schemes,
where the binary merge had an extra term of $O(\lg{\lg{K}})$ compared to the
multiway merge.
In addition, performing the merge operations in separate stages seems not to
have much effect on runtime, as may be expected due to the better exploitation
of locality in multiway merge.
Note that the binary merge scheme enables us to hide its runtime overhead, which
otherwise cannot be achieved via multiway merge.

We next compare how much memory the binary merge and the multiway merge needs
throughout the merge for the Sparse SUMMA algorithm.
For the multiway merge, this is simply given by the total number of elements that
are merged.
For the binary merge, since the merge operations are spread across stages, its
memory requirement is determined by the merge that contains the maximum number
of elements, which is typically the merge in the final stage.
Note that if the elements to be merged had all distinct keys, the binary and the
multiway merge would have the same memory overhead.
However, this is usually not the case and for that reason the binary merge leads
to significant memory savings.
Table~\ref{tb:merge-mem} displays the peak memory used by the binary merge and
the multiway merge in the first ten iterations of the MCL algorithm.
%
%
The binary merge has 20\%-25\% less memory overhead compared to the multiway
merge.
These figures show that apart from having roughly the same computational
overhead -which can effectively be hidden- with the multiway merge, the binary
merge is also more efficient in terms of memory usage.

\begin{table}[t]
  \centering
  \caption{Peak memory usage (GB) in first 10 iterations of the MCL algorithm
    (``mway'' denotes the multiway merge).}
  \scalebox{0.78} {
    \begin{tabular}{r r r r r r r r r r}
      \toprule
       & \multicolumn{3}{c}{\texttt{archaea}} & \multicolumn{3}{c}{\texttt{eukarya}} & \multicolumn{3}{c}{\isom} \\
      \cmidrule(r{4pt}){2-4} \cmidrule(r{4pt}){5-7} \cmidrule{8-10}
      MCL iter. & mway & binary & impr. & mway & binary & impr. & mway & binary & impr. \\
      \midrule      
      1  &  3.04 & 2.37 & 22\% & 6.50  & 5.11  & 21\% & 11.60 & 9.18  & 21\% \\     
      2  &  5.06 & 3.91 & 23\% & 13.56 & 10.63 & 22\% & 14.22 & 11.06 & 22\% \\  
      3  &  3.62 & 2.79 & 23\% & 8.37  & 6.54  & 22\% & 12.44 & 9.67  & 22\% \\     
      4  &  2.21 & 1.69 & 23\% & 4.20  & 3.24  & 23\% & 9.74  & 7.53  & 23\% \\      
      5  &  1.34 & 1.02 & 24\% & 2.37  & 1.80  & 24\% & 8.56  & 6.54  & 24\% \\      
      6  &  0.81 & 0.61 & 24\% & 1.36  & 1.03  & 24\% & 3.40  & 2.58  & 24\% \\      
      7  &  0.50 & 0.38 & 25\% & 0.75  & 0.57  & 25\% & 1.36  & 1.03  & 24\% \\      
      8  &  0.28 & 0.21 & 25\% & 0.37  & 0.28  & 25\% & 0.54  & 0.41  & 23\% \\      
      9  &  0.13 & 0.10 & 24\% & 0.15  & 0.12  & 23\% & 0.22  & 0.18  & 21\% \\      
      10 &  0.06 & 0.04 & 22\% & 0.06  & 0.05  & 20\% & 0.10  & 0.08  & 15\% \\      
    \bottomrule
    \end{tabular}%
    }
  \label{tb:merge-mem}%
    \vspace{3ex}
\end{table}%

\begin{figure*}[t]
  \centering
  \subfloat{\includegraphics[width=0.30\textwidth]{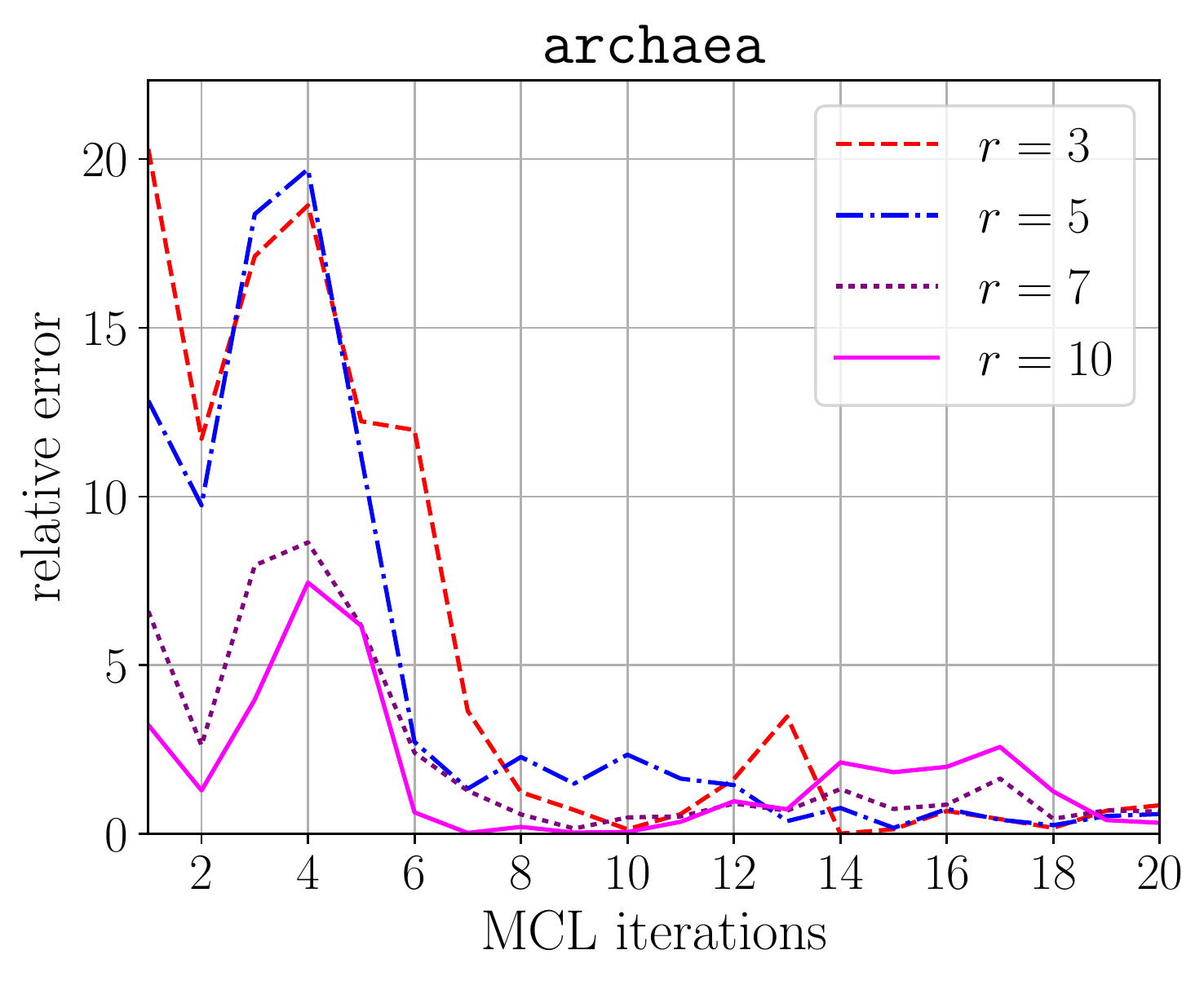}} 
  \subfloat{\includegraphics[width=0.30\textwidth]{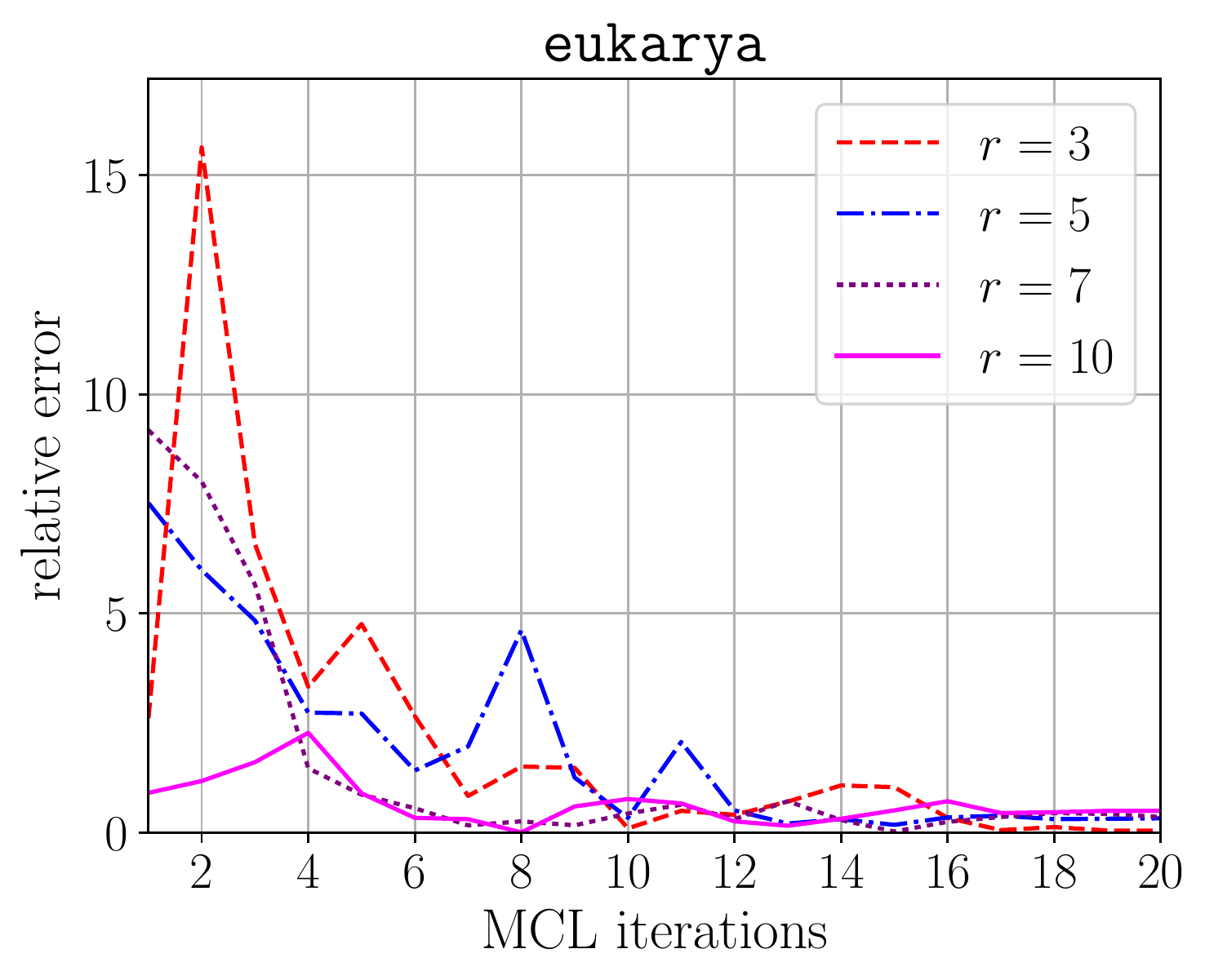}}
  \subfloat{\includegraphics[width=0.30\textwidth]{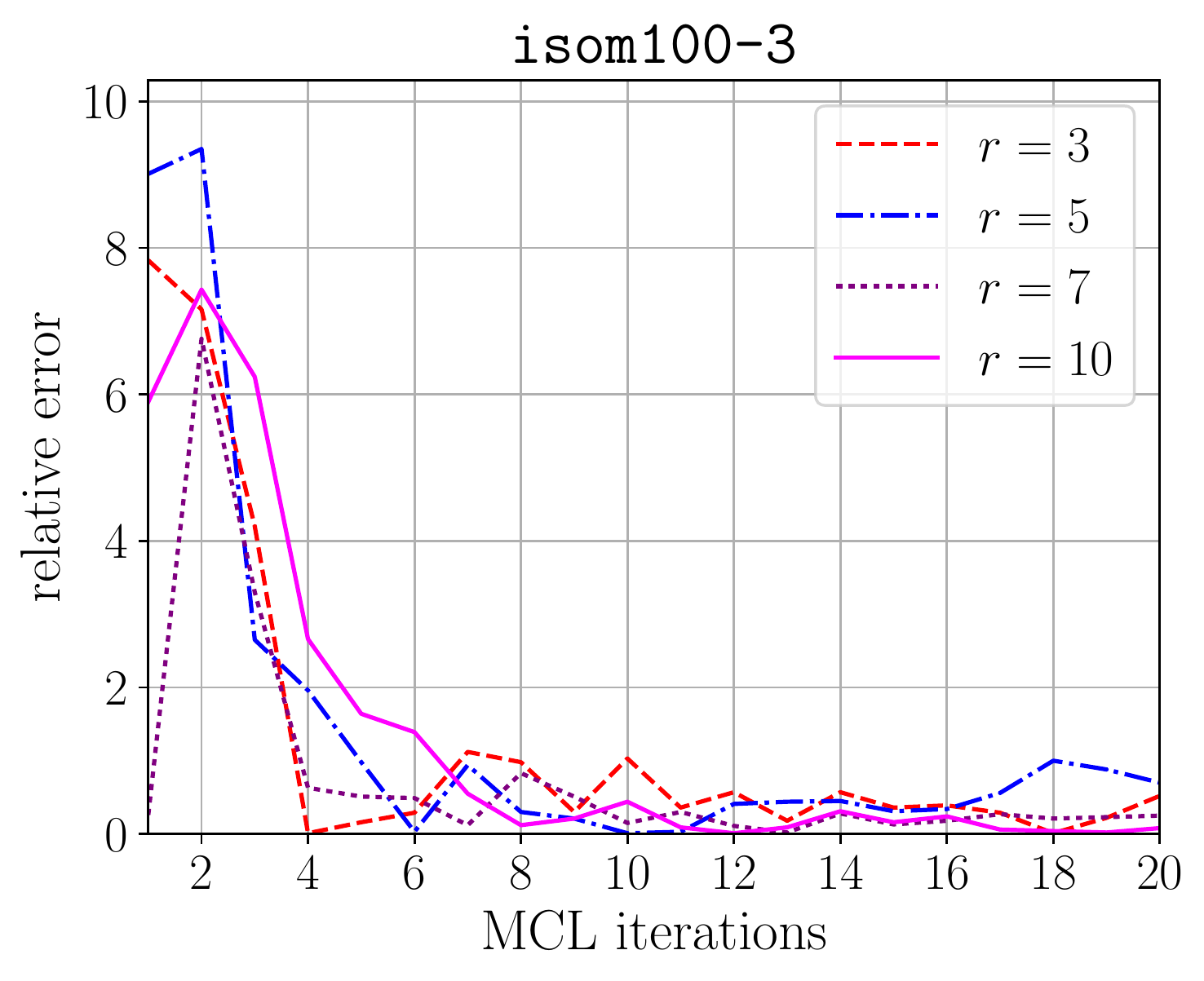}} \\ \vspace{-0.5em}
  \subfloat{\includegraphics[width=0.30\textwidth]{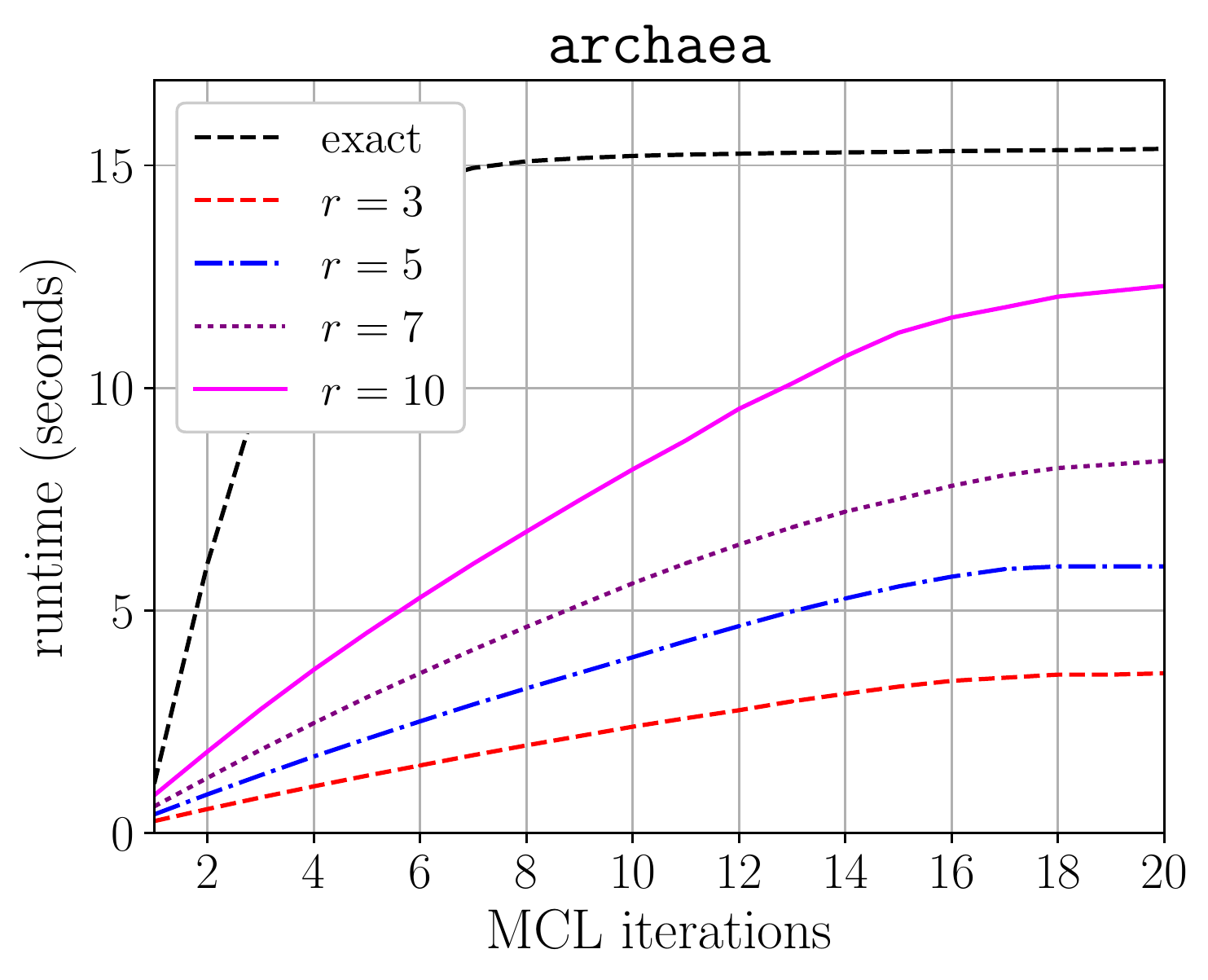}} 
  \subfloat{\includegraphics[width=0.30\textwidth]{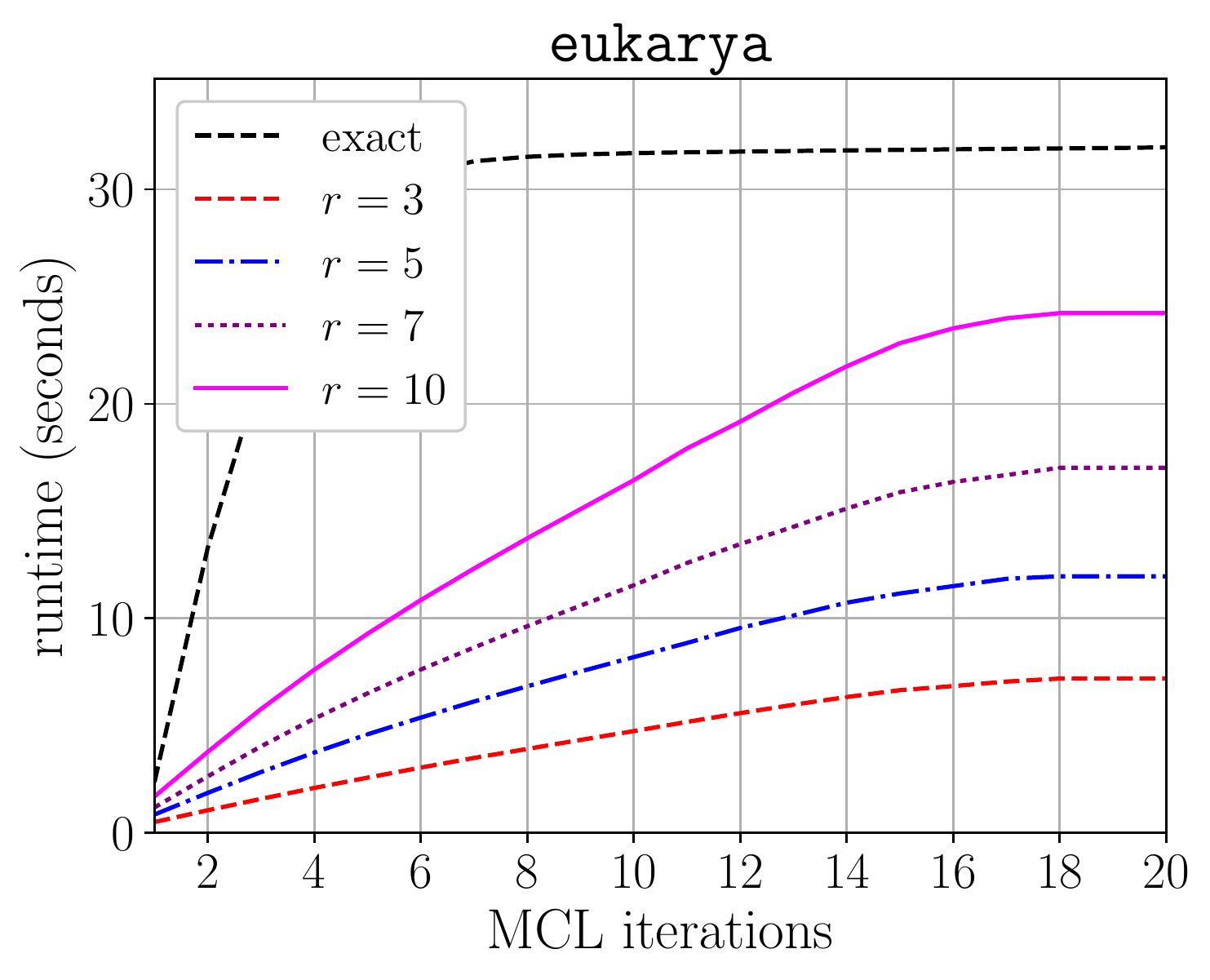}}
  \subfloat{\includegraphics[width=0.30\textwidth]{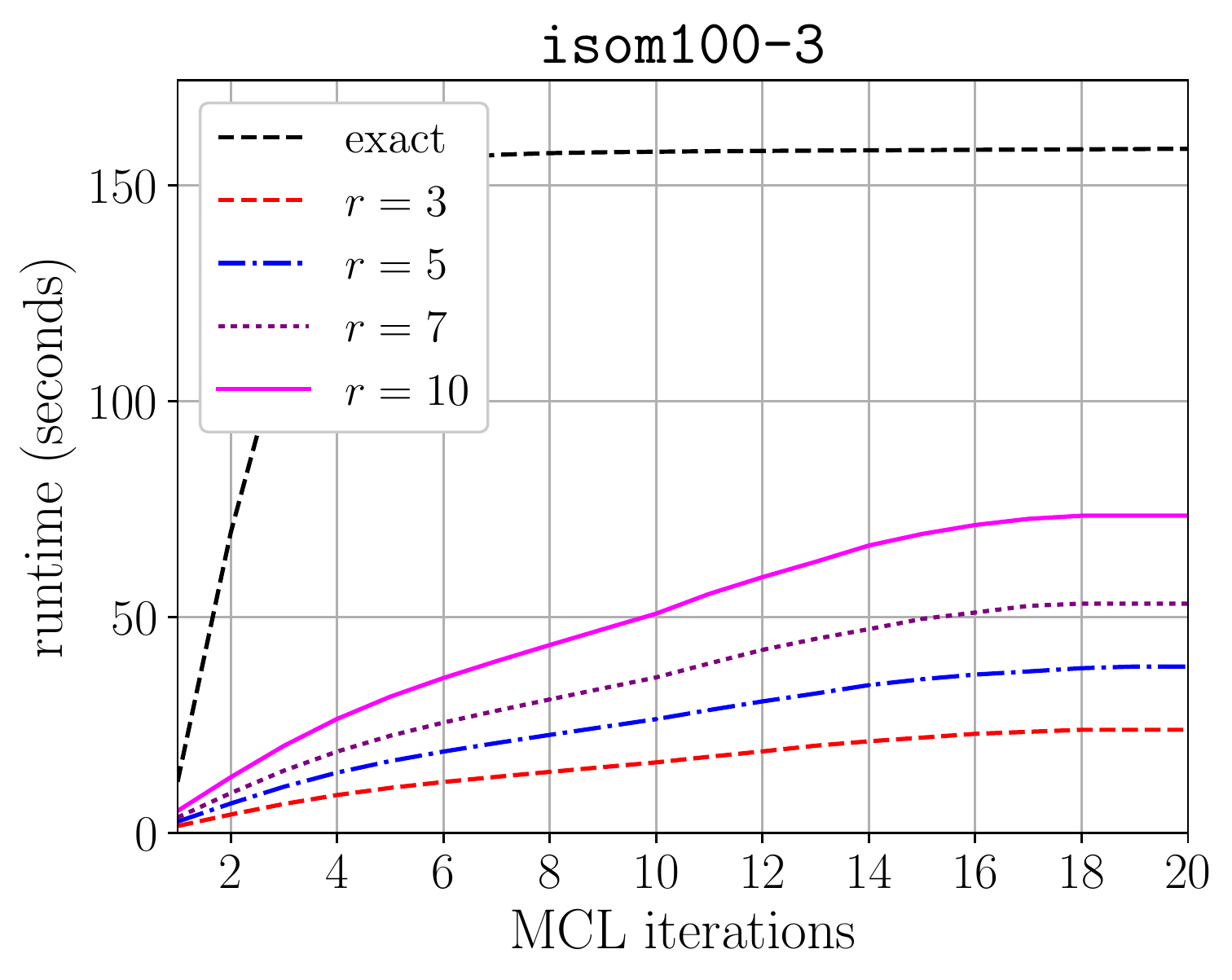}}
  \caption{Relative error and runtime of probabilistic memory requirement
    estimation.}
  \label{fig:est}
\end{figure*}

\subsection{Probabilistic memory requirement estimation}
\label{sec:exp-sampling}
We evaluate the performance of probabilistic memory requirement estimation in
terms of accuracy and runtime by comparing it to the exact memory requirement.
In probabilistic estimation, we use $\lambda = 1$ for the exponential
distribution and test out four different number of keys $r \in \{3, 5, 7, 10\}$.
The exact estimation is performed with the hash-table-based symbolic
multiplication.
Recall that the goal of both schemes is to compute the number of output elements
in local SpGEMM, which is then used to estimate the memory needed to run an
entire iteration of the MCL algorithm.
Both schemes make use of thread-level parallelism on a single node.
We use 40 threads per node for the experiments in this section and report the
averages obtained by the single MPI task at each node.
%

The top half of Figure~\ref{fig:est} illustrates the accuracy of the
probabilistic memory requirement estimation compared to the exact requirement.
%
%
By using relatively a few number of keys we can get an estimation that is within
10\% range of the correct estimation.
The worse prediction in the earlier iterations of the MCL algorithm can be
attributed to the columns having a higher variance in terms of nonzeros they
contain.
Compared to the number of phases estimated by the exact algorithm,
overestimation may lead to larger number of phases, while underestimation may
lead to smaller number of phases.
Underestimation is more critical as it can lead processes to go out of memory.
However, this can easily be compensated by providing a smaller value to \hmcl
than each process' actual available memory according to the relative error.

The bottom half of Figure~\ref{fig:est} compares the cumulative time spent in
memory requirement estimation for the exact and the probabilistic schemes.
The probabilistic scheme is faster compared to the exact scheme due to its
constant factor that does not change throughout the MCL iterations.
Recall that the number of operations performed by the probabilistic scheme is $r
\cdot \nnz(A) + r \cdot \nnz(B)$ and it is independent of the $\flops$ for $C=AB$.
%
For the exact scheme, however, the complexity depends on $\flops$, i.e., it is
$\compfac \cdot \nnz(C)$.
Therefore, a comparison between $\compfac$ and $r$ is partially helpful in explaining
the runtime differences between these two estimation schemes in the earlier
iterations (where the probabilistic scheme is faster) and the later iterations
(where the exact scheme is faster).
Relying on these observations, when $\compfac$ is below a certain threshold, we use
the exact scheme.



\begin{table}[t]
  \centering
  \caption{Runtime comparison of original HipMCL and the optimized
    HipMCL (``h'': hours, ``m'': minutes).}
  \scalebox{0.89} {
    \begin{tabular}{c r r}
      \toprule
      Network & \hmcl~\cite{Azad2018} & Optimized \hmcl \\
      \midrule
      \isomb               & \textbf{3.34h}, Summit, 100 nodes & \textbf{16.2m}, Summit, 100 nodes \\
      \midrule
      \multirow{2}{*}{\isomorig} & \multirow{2}{*}{\textbf{2.41h}, Cori, 2048 KNL nodes} & \textbf{22.6m}, Summit, 529 nodes  \\
                                 &                                                       & \textbf{14.1m}, Summit, 1024 nodes \\
      \midrule

      \texttt{metaclust50} & \textbf{3.23h}, Cori, 2048 KNL nodes & \textbf{1.04h}, Summit, 729 nodes \\
                                            
    \bottomrule
    \end{tabular}
    }
  \label{tb:bigruns}
  \vspace{3ex}
\end{table}

\subsection{Scalability and HipMCL on large instances}
Table~{\ref{tb:bigruns}} compares the overall runtime of original
HipMCL~{\cite{Azad2018}} and the HipMCL that is improved with the optimizations
in this work.
For the optimized HipMCL, we make use of both the CPU and GPU resources on a
node.
We run \texttt{isom100-1} on 100 nodes of Summit for both versions to see how much
improvement the optimized HipMCL obtains for a relatively large instance.
For \texttt{isom100-1}, the GPU-enabled and optimized HipMCL is $12.4\times$
faster than the original HipMCL.
We also run the optimized HipMCL for \texttt{isom100} and \texttt{metaclust50}
networks on Summit.
%
%
%
We did not run original HipMCL on these two networks on Summit because it
would take an extraordinary amount of compute hours due to original HipMCL not
utilizing the GPUs.
Instead, we present the clustering times reported in the original HipMCL paper
taken on the KNL nodes.
This is admittedly not an apples-to-apples comparison but it gives an idea how
much resource and time would be needed on a recent system with GPUs using the
optimized HipMCL, versus a relatively older system with no GPUs using the
original HipMCL.
The performance improvement we achieved relative to original HipMCL is more pronounced for \texttt{isom100} 
 compared to \texttt{metaclust50} because the SpGEMM runs on \texttt{isom100} have a 
 larger $\mathit{cf}$, leading to better utilization of GPUs.
%
%

\begin{figure}[t]
  \centering
  \subfloat{\includegraphics[width=0.25\textwidth]{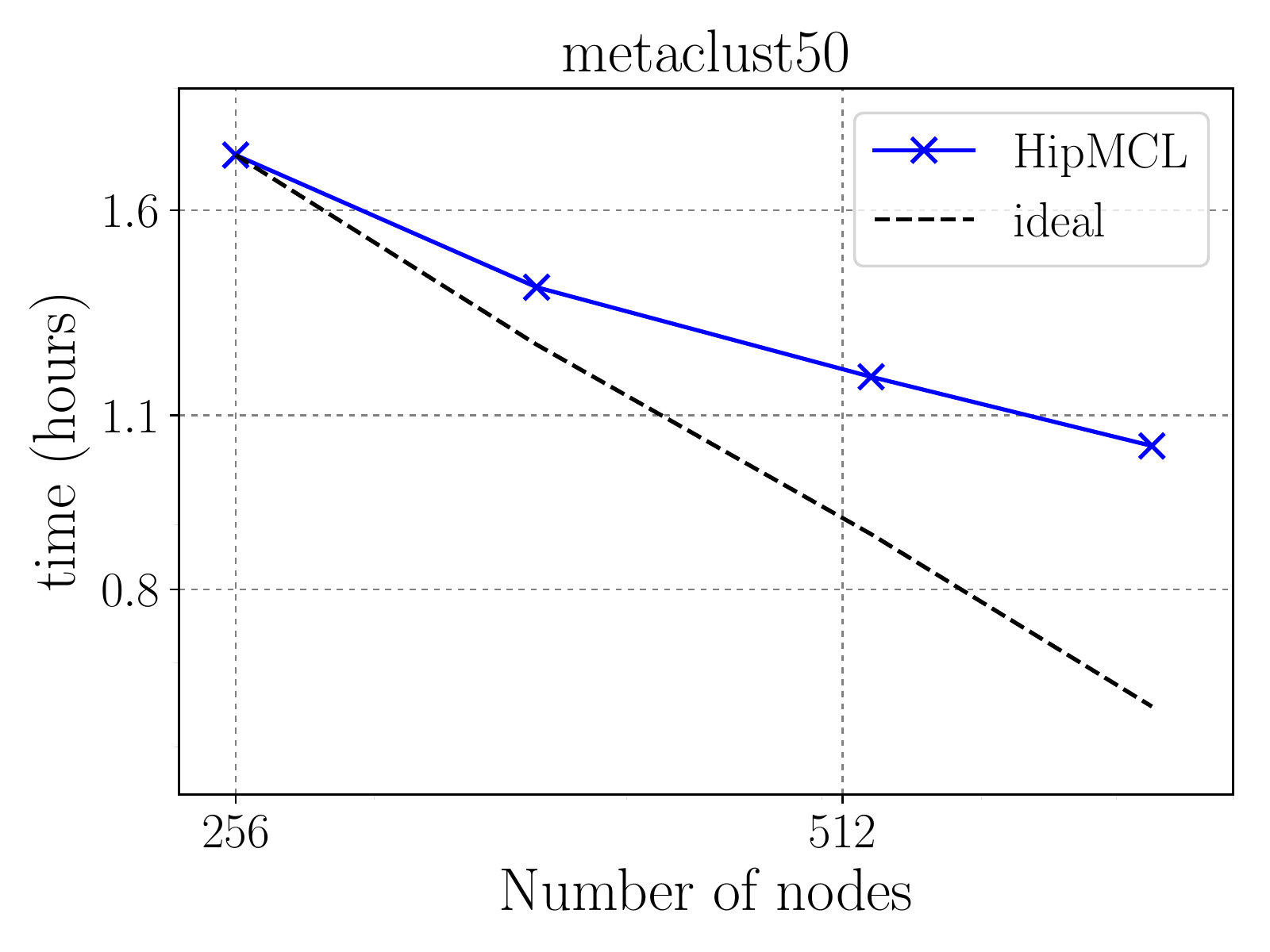}}
  \subfloat{\includegraphics[width=0.25\textwidth]{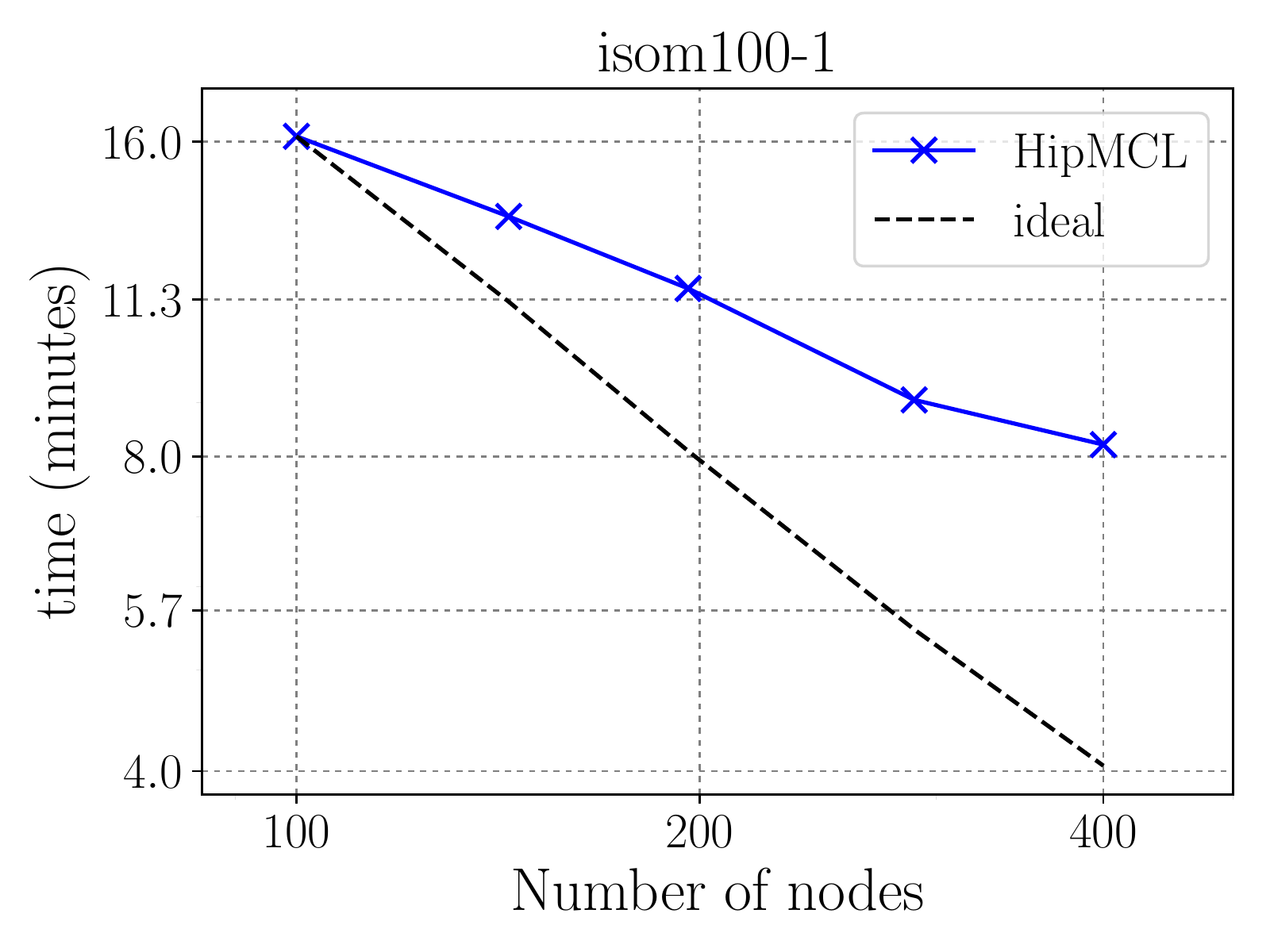}}   
  \caption{Strong scaling for \texttt{metaclust50} and \isomb.}
   \vspace{0.5ex}
  \label{fig:strong-scaling}
\end{figure}

Figure~\ref{fig:strong-scaling} presents the strong scaling plots of \isomb and
\texttt{metaclust50} networks.
The efficiency is 49\% and 57\% for \isomb and \texttt{metaclust50} networks,
respectively.

\begin{figure}[t]
  \centering
  \subfloat{\includegraphics[width=0.245\textwidth]{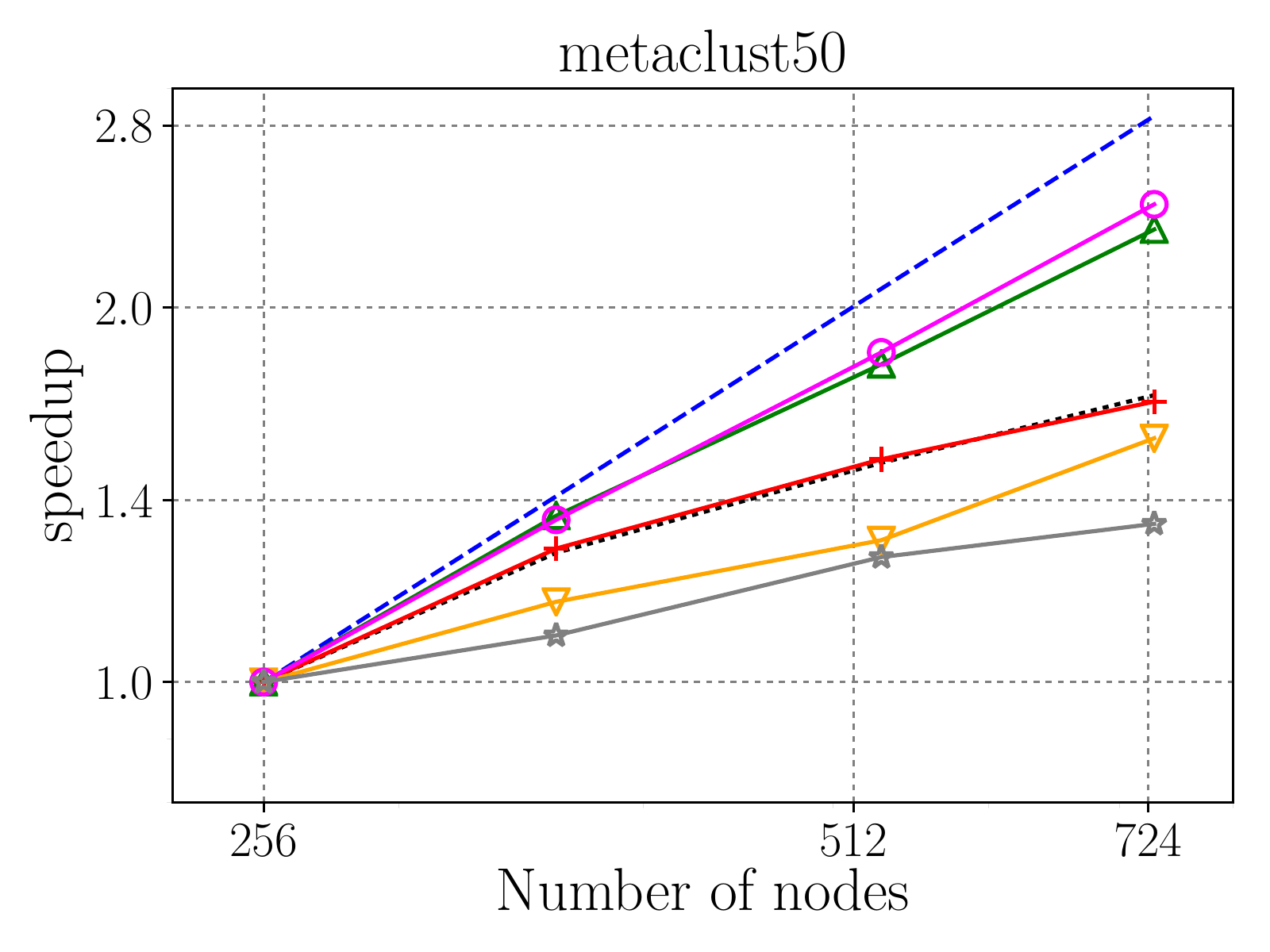}}
  \subfloat{\includegraphics[width=0.245\textwidth]{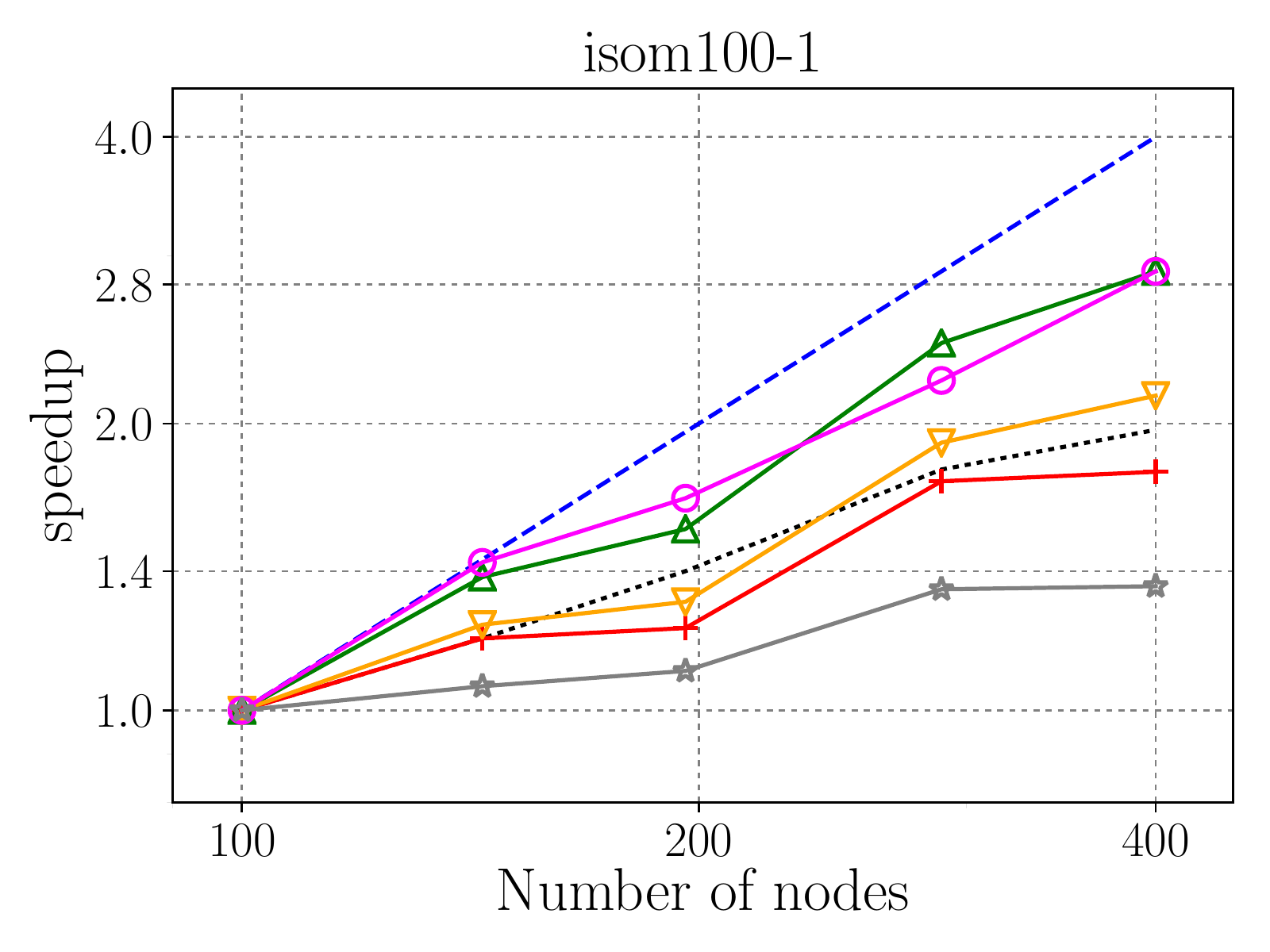}} \\
  \vspace{-2ex}
  \subfloat{\includegraphics[width=0.45\textwidth]{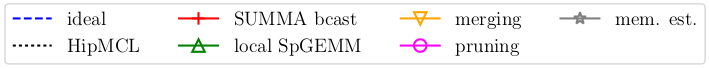}}
  \caption{Per-stage strong scaling analysis for \texttt{metaclust50} and \isomb.}
   \vspace{0.5ex}
  \label{fig:strong-scaling-analysis}
\end{figure}

Figure~\ref{fig:strong-scaling-analysis} further illustrates the strong scaling
behavior of different stages of optimized HipMCL with GPU support.
Note that the overall time is a complex combination of these stages; it is not
 equal to the sum of these stages because our Pipelined Sparse SUMMA algorithm
overlaps certain stages.
As seen from the figure, the biggest bottlenecks in scalability are the
memory requirement estimation, SUMMA broadcast, and merging.
The cost of the broadcast and merging operations can be hidden to a certain
extent by the Pipelined Sparse SUMMA.
Hence, the memory requirement estimation stage is a more
serious bottleneck: it takes roughly $2.5\times$ the SUMMA broadcast time
at 400 nodes for the \isomb network and $1.5\times$ the SUMMA broadcast time at 729 nodes for the \texttt{metaclust50} network.
The memory requirement estimation involves successive communication and
computational stages, as it mimics the execution of Sparse SUMMA algorithm.
In the future, we plan to port computations to the GPU and adapt a pipelined
methodology similar to the one used in Pipelined Sparse SUMMA to further optimize the
memory requirement estimation.

Finally, we provide the GPU and CPU idle times in the Pipelined Sparse SUMMA
algorithm with respect to varying number of nodes in Table~\ref{tb:idle-time}.
The CPU idle times usually tend to be higher than the GPU idles times.
The CPU waits for the GPU if the multiplication results are not yet computed,
and the GPU waits for the CPU if the broadcast operations are not yet completed.
The difference between the CPU and GPU idle times is larger in the \isomb
network because this network is denser than \texttt{metaclust50} and hence is more likely to be compute
intensive - as the CPU needs to stand more idle while the GPU is busy performing
the multiplication.
The GPU idle times can be reduced further, especially at large concurrencies,
via adapting 3D SpGEMM algorithm~\cite{Azad2016} in HipMCL.

\begin{table}[h]
  \centering
  \caption{CPU and GPU idle times in Pipelined Sparse SUMMA algorithm.}
  \scalebox{1.00} {
    \begin{tabular}{r r r r r r}
      \toprule
      \multicolumn{3}{c}{\isomb}  & \multicolumn{3}{c}{\texttt{metaclust50}}\\
      \cmidrule(r{4pt}){1-3} \cmidrule(r{4pt}){4-6} 
      & \multicolumn{2}{c}{idle time (sec.)}  & & \multicolumn{2}{c}{idle time (min.)} \\
      \cmidrule(r{4pt}){2-3} \cmidrule(r{4pt}){5-6}
      \#nodes & CPU & GPU  & \#nodes & CPU & GPU \\
      \midrule
      100 & 178.0 & 26.5 & 256 & 18.1 & 18.8 \\
      144 & 122.8 & 21.4 & 361 & 13.2 & 17.6 \\
      196 & 107.8 & 27.4 & 529 & 11.6 &  9.5 \\
      289 &  62.5 & 17.6 & 729 & 10.3 &  6.6 \\
      400 &  50.8 & 23.3 &     &      &      \\       
    \bottomrule
    \end{tabular}
    }
  \label{tb:idle-time}
\end{table}

\section{Conclusions}
\label{sec:conc}
Supercomputers are increasingly equipped with accelerators, especially
GPUs. Many complex applications are not able to take advantage of these
architectures due to a mismatch between the application demands and the hardware
constraints. In this work, we focused on such an application called HipMCL that
uses sparse data structures, performs significant amounts of communication, and
has challenging data access patterns. We demonstrated methods to optimize HipMCL
on large-scale supercomputers with GPU accelerators. Our methods include a new
merging algorithm that enable GPU-CPU computation pipelining, overlapping
communication with computation, probabilistic memory requirement estimation, and
integration of faster computational kernels. The resulting application is now
able to fully take advantage of the accelerators and the network. At scale, it
runs more than an order of magnitude faster for the same problem. This drastic
reduction in the time to solution enables clustering even larger biological
datasets such as those arising from metagenomic studies when the data sizes have
been increasing at a rate much faster than Moore's law.

\section*{Acknowledgments}

This research was supported in part by the Applied Mathematics program of the DOE Office of Advanced Scientific Computing
Research under Contract No. DE-AC02-05CH11231, and in part by the
Exascale Computing Project (17-SC-20-SC), a collaborative effort of
the U.S. Department of Energy Office of Science and the National
Nuclear Security Administration.

We used resources of the NERSC supported by the Office of Science of the DOE under Contract No. DEAC02-05CH11231. 
This research also used resources of the Oak Ridge Leadership Computing Facility at the Oak Ridge National Laboratory, which is supported by the Office of Science of the U.S. Department of Energy under Contract No. DE-AC05-00OR22725.





%

%
\bibliographystyle{IEEEtran}
\bibliography{oguzall}

\end{document}